\documentclass[superscriptaddress,aps,preprintnumbers,amsmath,amssymb,prd,nofootinbib,preprint]{revtex4-1}
\pdfoutput=1
\usepackage{graphicx}
\usepackage{epstopdf}
\usepackage{dcolumn}% Align table columns on decimal point
\usepackage{bm}% bold math
\usepackage{bbm}
\usepackage{hyperref}
\usepackage[usenames, dvipsnames]{color}
\usepackage{amsmath,amssymb}
\usepackage[sort&compress]{natbib}
\usepackage{float}
\usepackage{multirow}
\usepackage{slashed}
\usepackage{cancel}
\usepackage[table]{xcolor}
\usepackage{booktabs}
\usepackage{makecell}
\usepackage{mathtools}
\usepackage{pifont}
\usepackage{enumitem}
\usepackage{ulem}

\def\mpl{M_{\rm Pl}}

\numberwithin{equation}{section}

\makeatletter
\renewcommand{\p@subsection}{}
\renewcommand{\p@subsubsection}{}
\makeatother

\makeatletter
\def\l@subsubsection#1#2{}
\makeatother

\makeatletter
    \def\CT@@do@color{%
      \global\let\CT@do@color\relax
            \@tempdima\wd\z@
            \advance\@tempdima\@tempdimb
            \advance\@tempdima\@tempdimc
    \advance\@tempdimb\tabcolsep
    \advance\@tempdimc\tabcolsep
    \advance\@tempdima2\tabcolsep
            \kern-\@tempdimb
            \leaders\vrule
                    \hskip\@tempdima\@plus  1fill
            \kern-\@tempdimc
            \hskip-\wd\z@ \@plus -1fill }
    \makeatother
 
\makeatletter
\def\@ssect@ltx#1#2#3#4#5#6[#7]#8{%
  \def\H@svsec{\phantomsection}%
  \@tempskipa #5\relax
  \@ifdim{\@tempskipa>\z@}{%
    \begingroup
      \interlinepenalty \@M
      #6{%
       \@ifundefined{@hangfroms@#1}{\@hang@froms}{\csname @hangfroms@#1\endcsname}%
       {\hskip#3\relax\H@svsec}{#8}%
      }%
      \@@par
    \endgroup
    \@ifundefined{#1smark}{\@gobble}{\csname #1smark\endcsname}{#7}%
  }{%
    \def\@svsechd{%
      #6{%
       \@ifundefined{@runin@tos@#1}{\@runin@tos}{\csname @runin@tos@#1\endcsname}%
       {\hskip#3\relax\H@svsec}{#8}%
      }%
      \@ifundefined{#1smark}{\@gobble}{\csname #1smark\endcsname}{#7}%
      \addcontentsline{toc}{#1}{\protect\numberline{}#8}%
    }%
  }%
  \@xsect{#5}%
}%
\makeatother

\hypersetup{colorlinks,citecolor= blue,linkcolor= black}

\begin{document}

%%%%%%%%%%%%%%%%%%%%%%%%%%%%%%%%%%%%%%%%%%%

\newcommand{\meV}{ \ {\rm meV} }
\newcommand{\eV}{ \ {\rm eV} }
\newcommand{\keV}{ \ {\rm keV} }
\newcommand{\MeV}{\  {\rm MeV} }
\newcommand{\GeV}{\  {\rm GeV} }
\newcommand{\TeV}{\  {\rm TeV} }
\newcommand{\PeV}{\  {\rm PeV} }
\newcommand{\EeV}{\  {\rm EeV} }

%domain wall number
\newcommand{\ndw}{N_{\rm DW}}

%scale factor%
\newcommand{\scale}{R}

\newcommand{\fluid}{{\rm rot}}

%physical momentum%
\newcommand{\kp}{{k_{p}}}

\newcommand{\KFM}{KFM}

\newcommand{\paren}[1]{\left(#1\right)} %parenthesis
\newcommand{\fig}[1]{Fig.~\ref{#1}}
\newcommand{\eq}[1]{Eq.(\ref{#1})}

\title{
Acoustic Misalignment Mechanism for Axion Dark Matter
}

\preprint{FERMILAB-PUB-25-0110-V}

\author{Arushi Bodas}
\affiliation{Department of Physics, University of Chicago, Chicago, IL 60637, USA}
\affiliation{Enrico Fermi Institute and Kavli Institute for Cosmological Physics, University of Chicago, Chicago, IL 60637, USA}
\affiliation{Particle Theory Department, Fermilab, Batavia, Illinois 60510, USA}

\author{Raymond T. Co}
\affiliation{Physics Department, Indiana University, Bloomington, IN, 47405, USA}

\author{Akshay Ghalsasi}
\affiliation{Jefferson Physical Laboratory, Harvard University, Cambridge, MA 02138, USA}

\author{Keisuke Harigaya}
\affiliation{Department of Physics, University of Chicago, Chicago, IL 60637, USA}
\affiliation{Enrico Fermi Institute and Kavli Institute for Cosmological Physics, University of Chicago, Chicago, IL 60637, USA}
\affiliation{Kavli Institute for the Physics and Mathematics of the Universe (WPI),\\
        The University of Tokyo Institutes for Advanced Study,\\
        The University of Tokyo, Kashiwa, Chiba 277-8583, Japan}

\author{Lian-Tao Wang}
\affiliation{Department of Physics, University of Chicago, Chicago, IL 60637, USA}
\affiliation{Enrico Fermi Institute and Kavli Institute for Cosmological Physics, University of Chicago, Chicago, IL 60637, USA}

\begin{abstract}
A rotation in the field space of a complex scalar field corresponds to a Bose-Einstein condensation of $U(1)$ charges. We point out that fluctuations in this rotating condensate exhibit sound-wave modes, which can be excited by cosmic perturbations and identified with axion fluctuations once the $U(1)$ charge condensate has been sufficiently diluted by cosmic expansion. We consider the possibility that these axion fluctuations constitute dark matter and develop a formalism to compute its abundance. We carefully account for the growth of fluctuations during the epoch where the complex scalar field rotates on the body of the potential and possible nonlinear evolution when the fluctuations become non-relativistic. We find that the resultant dark matter abundance can exceed the conventional and kinetic misalignment contributions if the radial direction of the complex scalar field is sufficiently heavy. The axion dark matter may also be warm enough to leave imprints on structure formation. We discuss the implications of this novel dark matter production mechanism---{\it acoustic misalignment mechanism}---for the axion rotation cosmology, including kination domination and baryogenesis from axion rotation, as well as for axion searches. 
\end{abstract}

\maketitle

\begingroup
\hypersetup{linkcolor=black}
\renewcommand{\baselinestretch}{1.12}\normalsize
\tableofcontents
\renewcommand{\baselinestretch}{2}\normalsize
\endgroup

%%%%%%%%%%%%%%%%%%%%%%%%%%%
\section{Introduction}
\label{sec:intro}
%%%%%%%%%%%%%%%%%%%%%%%%%%%%

Spontaneously broken symmetries play important roles in the solutions to a variety of problems in the Standard Model (SM). 
Examples of such symmetries include Peccei-Quinn (PQ) symmetry~\cite{Peccei:1977hh,Peccei:1977ur} (a solution to the strong CP problem), lepton symmetry~\cite{Chikashige:1980ui} (the origin of neutrino masses), and flavor symmetry~\cite{Froggatt:1978nt} (the origin of the pattern of the fermion masses and mixings). 
If the symmetry is global, the spontaneous breaking is associated with a Nambu-Goldstone boson (NGB). 
If the symmetry is explicitly broken, the NGB obtains a nonzero mass. 
We refer to these pseudo Nambu-Goldstone bosons generically as ``axions."

The axion is the angular direction of the complex scalar field whose radial vacuum expectation value (VEV) spontaneously breaks the global symmetry, which we call the PQ symmetry. The complex scalar field, which we call the PQ field in this paper, may rotate in its field space in the early universe, initiated by the Affleck-Dine mechanism~\cite{Affleck:1984fy}. The rotation corresponds to a nonzero charge associated with the global symmetry, which we call the PQ charge or PQ asymmetry. 
If the PQ charge density is large enough, the rotation is stable against dissipation into thermal fluctuations, because it is free-energetically favorable to keep the charges in the form of rotation, i.e., a Bose-Einstein condensate, rather than in the form of a particle-antiparticle asymmetry in the bath~\cite{Co:2019wyp,Domcke:2022wpb}.

It is known that the axion rotation can explain both the dark matter abundance and the matter-antimatter asymmetry of the universe.
The kinetic energy associated with the axion rotation may be converted into the axion dark matter abundance in the form of a coherent axion oscillation~\cite{Co:2019jts} or axion fluctuations~\cite{Co:2021rhi,Eroncel:2022vjg} (see also~\cite{Jaeckel:2016qjp,Berges:2019dgr}).
This is known as the kinetic misalignment mechanism (KMM).
In the early stages of the axion rotation, the radial-mode oscillation can also produce axion dark matter via parametric resonance~\cite{Co:2017mop,Co:2020dya}.
The matter-antimatter asymmetry can be explained if a part of the PQ charge is transferred into baryon asymmetry~\cite{Co:2019wyp,Domcke:2020kcp,Co:2020xlh,Co:2020jtv,Harigaya:2021txz,Chakraborty:2021fkp,Kawamura:2021xpu,Co:2021qgl,Co:2022aav,Barnes:2022ren,Co:2022kul,Berbig:2023uzs,Chun:2023eqc,Barnes:2024jap,Wada:2024cbe,Datta:2024xhg}, which is known as the axiogenesis mechanism.

The axion rotation follows an interesting equation of state. In its initial phase, when the PQ field rotates on the body of the potential, the rotation behaves as matter or radiation, depending on whether the potential of the PQ field is nearly quadratic or quartic. We will call this phase the pre-kination phase. After the field reaches the bottom of the potential, the rotation behaves as kination~\cite{Co:2019wyp}, henceforth referred to as the kination phase.

\begin{figure}
    \centering
    \includegraphics[width=0.8\linewidth]{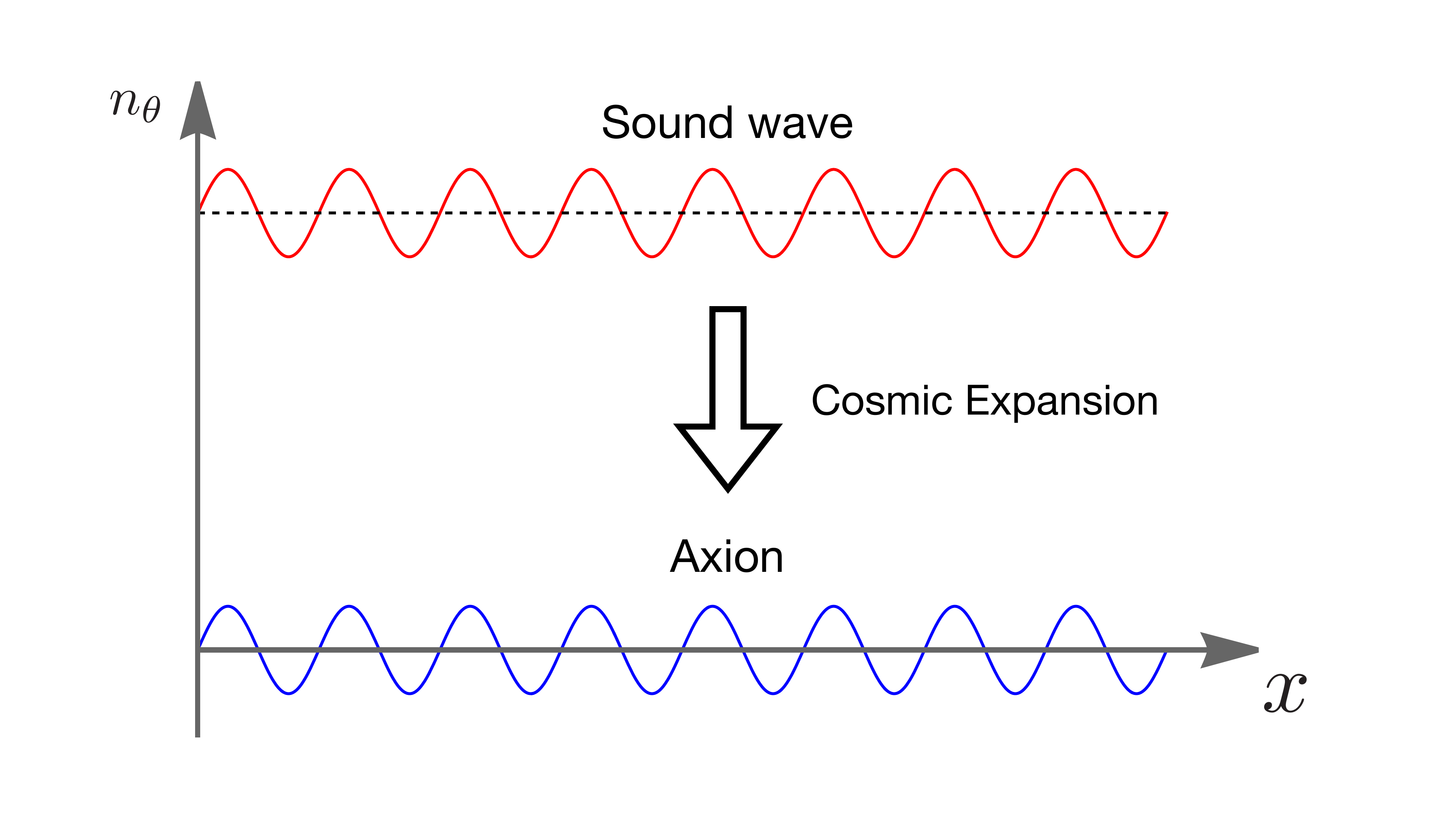}
    \caption{Sound waves of the PQ charge density $n_\theta$ are produced by cosmic perturbations. After the dilution of the charge density by cosmic expansion,  sound waves become axions. The dotted line shows the average charge density.}
    \label{fig:sound wave-axion}
\end{figure}

Like any other energy component of the universe, the axion rotation would have density perturbations, which may be sourced by adiabatic perturbations that explain the observed anisotropies in the cosmic microwave background, or are simply isocurvature perturbations of the rotation.
Since the PQ field is complex, the perturbations have two modes. We point out that one of them is a phonon mode, i.e., sound waves of the PQ charge density, and may be identified with axion fluctuations once the PQ charge density is diluted by cosmic expansion.
The schematic picture is shown in Fig.~\ref{fig:sound wave-axion}.
Initially, the PQ charge density $n_\theta$ is large and there is a NGB mode associated with the spontaneous breaking of the PQ symmetry and the time-translational symmetry into a diagonal subgroup, which is the sound-wave mode. Sound waves of $n_\theta$ can be produced by primordial perturbations. 
After $n_\theta$ decreases due to cosmic expansion, since the PQ symmetry is spontaneously broken also at the vacuum, the sound waves smoothly become NGBs at the vacuum, i.e., axions.

The computation of the perturbations of the axion rotation has been developed in~\cite{Co:2021lkc,Harigaya:2023mhl,Co:2024oek} for the era when the PQ field rotates at the body and bottom of its potential,
and in~\cite{Eroncel:2022vjg,Eroncel:2024rpe,Eroncel:2025bcb} for the bottom of potential.
In~\cite{Eroncel:2025bcb}, it was pointed out that the second-order perturbations of kination fluid behave as radiation and clarified why we may compute the radiation abundance by the linear perturbation theory despite the second-order energy perturbations eventually exceeding the zeroth-order kination energy.
Ref.~\cite{Eroncel:2025bcb} also noted the possibility that the radiation produced from cosmic perturbations could become dark matter.
However, the resultant dark matter abundance was not computed and the evolution of the perturbations in the pre-kination phase was not considered.
Ref.~\cite{Co:2020dya} discussed parametric resonance production of axion fluctuations from the radial excitation on the rotating background rather than the production by primordial cosmic perturbations. 

\begin{figure}
    \centering
    \includegraphics[width=0.8\linewidth]{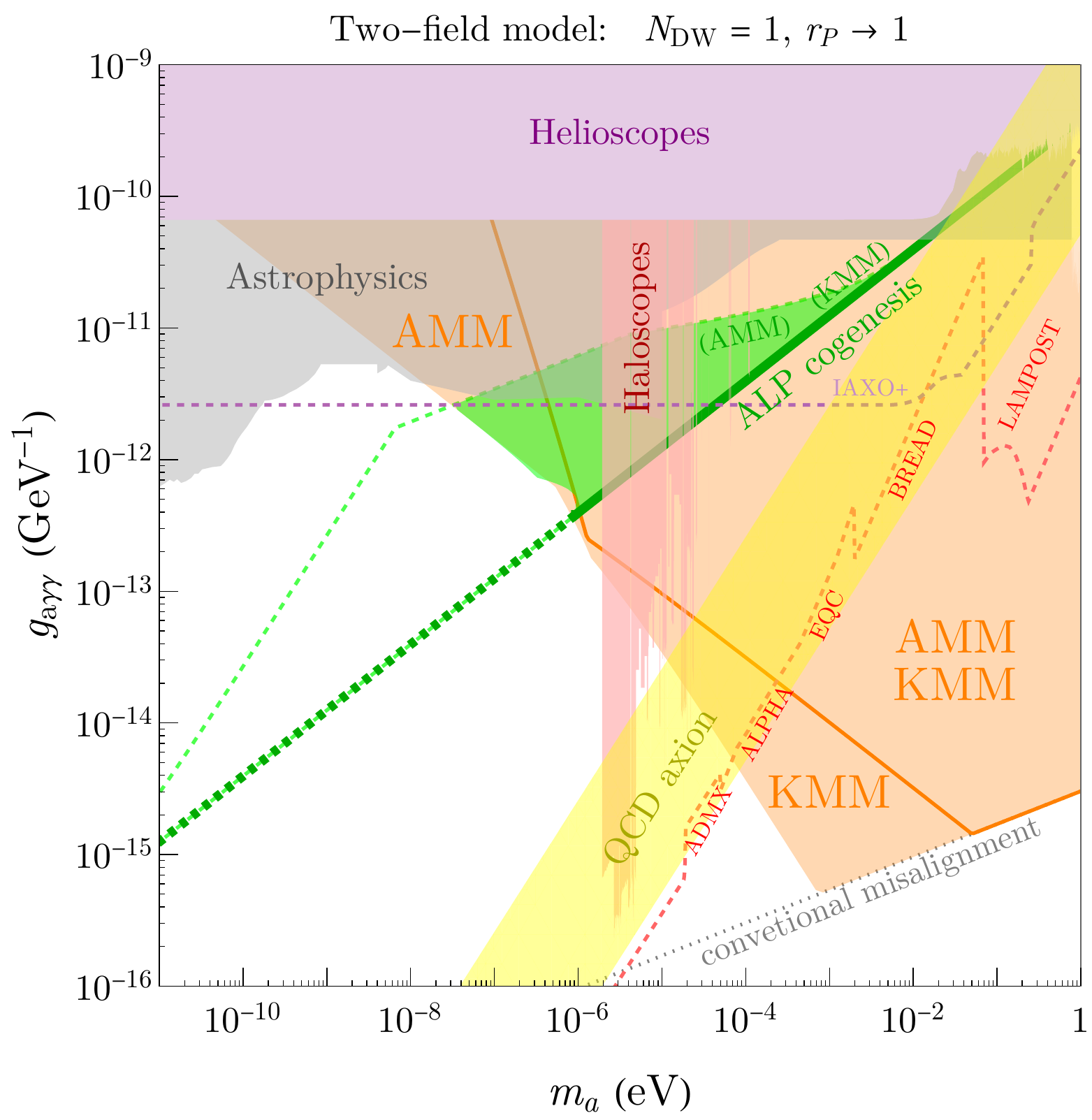}
    \caption{The acoustic misalignment mechanism (AMM) and the kinetic misalignment mechanism (KMM) can explain the observed dark matter abundance in the orange-shaded region, with the labels showing which mechanism can dominate.  In the green-shaded region and on the sold green line, the observed baryon asymmetry can also be explained by axiogenesis via the weak sphaleron process. Further details are provided in Secs~\ref{sec:implications} and~\ref{sec:axiogenesis}.}
    \label{fig:AMM_cogen_rP1}
\end{figure}

In this paper, we compute the axion dark matter abundance resulting from the cosmic perturbations of the sound-wave mode of the axion rotation, which we call the {\it Acoustic Misalignment Mechanism} (AMM). 
We take into account the evolution of fluctuations during the pre-kination phase, which is important for the computation of the contribution from the modes that enter the horizon during the pre-kination era. We find that the contribution from those modes may dominate over the contribution from the modes that enter the horizon during the kination phase if the axion rotation dominates the universe. Rotation domination is necessary in certain regions of the parameter space in order to explain the observed dark matter abundance.
We discover possible nonlinear evolution if the fluctuations are too large when they become non-relativistic.
We compute the axion abundance from the AMM as a function of the model parameters and find that the AMM contribution dominates over the KMM contribution when the PQ field mass is sufficiently large. 
We also find that axion dark matter produced by the AMM may be warm enough to affect the structure formation of the universe.
These results have implications for the axiogenesis scenarios as well.

Fig.~\ref{fig:AMM_cogen_rP1} summarizes the implications for the axion searches. The AMM and KMM can explain the observed dark matter abundance in the orange-shaded region with the labels showing which mechanism dominates.
We see that the AMM and KMM predict larger axion couplings than the conventional misalignment mechanism~\cite{Preskill:1982cy,Abbott:1982af,Dine:1982ah}, shown by the gray dotted line with the misalignment angle assumed to be unity. 
Inside the green-shaded region (along the solid green line), the observed baryon asymmetry of the universe can be explained by axiogenesis using the electroweak sphaleron process, along with the dark matter produced by the AMM (KMM) mechanism. Current constraints and projections for future axion searches are taken from~\cite{AxionLimits}.

This paper is structured as follows.
Sec.~\ref{sec:rotation} reviews the zero-mode evolution, the equation of state of the axion rotation, and the constraints from the efficient thermalization of rotation. 
In Sec.~\ref{sec:fluctuations}, we analyze the evolution of the cosmic perturbations of the axion rotation.
The results are then used in Sec.~\ref{sec:DM} to compute the axion abundance from the AMM, which is compared to the KMM contribution. Sec.~\ref{sec:DM} also discusses implications for the axiogenesis scenarios. Sec.~\ref{sec:summary} provides a summary of our findings and discussion. Appendix~\ref{app:phonon} derives the dispersion relation of two perturbation modes around the rotating background and explicitly shows the existence of sound-wave modes. Appendix~\ref{sec:extra figs} provides extra figures.

%%%%%%%%%%%%%%%%%%%%%%%%%%%
\section{Zero-mode Evolution of Axion Rotation}
\label{sec:rotation}
%%%%%%%%%%%%%%%%%%%%%%%%%%%%

In this section, we describe the evolution of the zero mode of the axion rotation. We discuss how the rotation is initiated and becomes circular by thermalization, and how the equation of state of the rotation evolves. 

The axion field $a $ is the angular direction of a complex scalar field $P$, which can be decomposed as
\begin{equation}
    P = \frac{1}{\sqrt{2}} r e^{i \theta} = \frac{1}{\sqrt{2}} r e^{i \theta_a/\ndw}.
\end{equation}
The radial mode obtains a VEV of $r = f_a \ndw$ with $f_a$ the decay constant and $\ndw$ the domain wall number.
In the first equality, we decomposed $P$ simply into the radial component $r$ and the angular component $\theta$. In the second equality, we normalized the angular variable $\theta_a = a/f_a$ so that the periodicity of its dominant vacuum potential $\propto \cos(\ndw \, \theta)$ is $2\pi$.

\subsection{Initiation of rotation}

Throughout this work, we focus on the case where the complex field $P$ undergoes a rotational motion in the field space. 
This dynamics is well-motivated in the early universe and can be triggered as follows. 
As pointed out in the Affleck-Dine baryogenesis mechanism~\cite{Affleck:1984fy}, if a complex scalar field starts with a large displacement from the origin, higher-dimensional operators that explicitly break the $U(1)$ symmetry can be sizable and generate a ``kick'' in the angular direction, initiating a rotation, when the Hubble scale drops below the mass of the radial mode. 
The large initial field value can be a result of a mere initial condition during inflation or of a negative Hubble-induced mass during inflation. See~\cite{Dine:1995uk,Dine:1995kz,Harigaya:2015hha} for the detail of the dynamics involving a Hubble~induced mass. Once the rotation begins, the radius of the motion decreases due to redshift and the higher-dimensional operators become suppressed and irrelevant. The rotation then corresponds to a nonzero conserved $U(1)$ charge, whose yield is given by
\begin{equation}
    Y_\theta = \frac{n_\theta}{s} =
    \begin{cases}
            \frac{\dot{\theta}_af_a^2}{s}
 & r \simeq \ndw f_a \\
 \frac{m_r r^2/\ndw}{s} & r \gg \ndw f_a,
    \end{cases}
\end{equation}
where $s$ is entropy density, $\dot\theta_a = {\rm d}\theta_a/{\rm d}t$ is the angular velocity, and $m_r(r)$ is the local curvature (mass) of the potential of the radial direction.

For the initial field value $r_i$ at the start of rotation, $Y_\theta$ is given by
\begin{equation}
    Y_\theta \simeq \epsilon \left. \frac{m_r(r_i) r_i^2}{ \ndw s }\right|_{m_r(r_i)= 3H} \simeq 600\times \frac{\epsilon}{\ndw} \left(\frac{r_i}{10^{16}{\rm GeV}}\right)^2 \left(\frac{\rm TeV}{m_r(r_i)}\right)^{1/2}\left(\frac{106.75}{g_*}\right)^{1/4},
\end{equation}
where $\epsilon \sim (\partial V/\partial\theta)/(r \partial V/\partial r)$ parametrizes the strength of the kick and $g_*$ captures effective degrees of freedom of the thermal bath.
As we will see in Sec.~\ref{sec:DM}, for the rotation to explain the observed dark matter abundance, $Y_\theta\gg 1$ is typically required.
To obtain such large $Y_\theta$, we need $m_r(r_i) \ll r_i$ at large $r_i$, which requires a flat potential of $r$.
For the model with a positive quartic and negative quadratic term in the potential, a small quartic coupling is required.
In supersymmetric theories, a flat potential can be naturally obtained. 
In fact, the converse of the theorem in~\cite{Affleck:1983vc} tells us that in supersymmetric theories, as long as the PQ-breaking sector does not spontaneously break supersymmetry, there should be a flat direction that corresponds to the extension of the $U(1)$ symmetry transformation parameter into the complex plane. Couplings with supersymmetry breaking lift the flatness, but as long as the PQ symmetry breaking scale is much larger than the soft mass scale, the flatness is lifted only by the soft supersymmetry-breaking mass of $P$.  
Two types of supersymmetric models are introduced in the next subsection.

When the motion is initiated, the $P$ field receives a kick not only in the angular direction but also in the radial direction. 
This implies that the motion is initially elliptical, and there is energy associated with the radial-mode oscillations. 
To avoid a moduli problem, the radial mode $r$ should dissipate via interactions with the thermal bath.
After thermalization,  $U(1)$ charge conservation dictates that $P$ must follow the motion with the least energy for a fixed charge, which is circular. While a part of the $U(1)$ charge can be transferred into particle-antiparticle asymmetry in the thermal bath, it is still free-energetically favored to keep the majority of charges in the rotation as long as $Y_\theta \gg m_r /T$~\cite{Co:2019wyp}.

\subsection{Equation of state of rotation}

The evolution of the energy density of the circular rotation depends on the form of the radial potential. 
For a model with a negative quadratic term and a positive quartic term (referred to as the ``quartic model'' henceforth),
\begin{equation}
    V(P) = \lambda \left(|P|^2 - \frac{1}{2}\ndw^2f_a^2\right)^2,
\end{equation}
the potential is nearly quartic for $r \gg \ndw f_a $. 
The energy of the rotation redshifts as radiation, $\rho_\theta \propto R^{-4}$,  where $R$ is the scale factor of the universe. Once $r$ reaches the minimum $\ndw f_a$, the energy of the rotation is dominated by the kinetic energy and hence redshifts as kination, $\rho_\theta \propto R^{-6}$.

In supersymmetric theories, $V(P)$ at $r \gg \ndw f_a$ can be nearly quadratic.
One supersymmetric model is the PQ breaking by dimension transmutation with the potential~\cite{Moxhay:1984am} 
\begin{align}
\label{eq:one-field}
V(P) = \frac{1}{2} m_r^2 |P|^2 \left( {\rm ln} \frac{2 |P|^2}{f_a^2 N_{\rm DW}^2} -1 \right),
\end{align}
where $m_r$ is the soft mass of the radial mode, and the logarithmic factor arises from a radiative correction from a Yukawa coupling of $P$ with PQ-charged fields, such as the heavy Kim-Shifman-Vainshtein-Zakharov~\cite{Kim:1979if,Shifman:1979if} (KSVZ) quarks. We call this model ``the log-potential model."

Another type of model is the PQ breaking by two PQ-charged complex fields, where the superpotential and the soft supersymmetry-breaking potential read
\begin{align}
\label{eq:W_two-field}
    W = \lambda X (P \bar{P} - v_{\rm PQ}^2),~~V_{\rm soft}(P) = m_P^2 |P|^2 + m_{\bar{P}}^2|\bar{P}|^2. 
\end{align}
The $F$-term of the chiral multiplet $X$ generates a moduli space along which $P \bar P = v_{\rm PQ}^2$ and the PQ symmetry is spontaneously broken. We identify $P$ as the one that takes the large initial field value we assume in this work. Then $\bar P$ can be integrated out by replacing $\bar P = v_{\rm PQ}^2/P$, giving an effective potential of $P$ as 
\begin{equation}
\label{eq:two-field}
    V(P) \simeq m_P^2 |P^2| \left( 1 + r_P^2 \frac{v_{PQ}^4}{|P^4|} \right), 
\end{equation}
with $r_P \equiv m_{\bar{P}}/m_P$ the ratio of the soft masses of the two fields. We call this model ``the two-field model."

\begin{figure}[t!]
	\centering	\includegraphics[width=0.75\columnwidth,trim={3.5cm 4.5cm 5cm 4.5cm},clip]{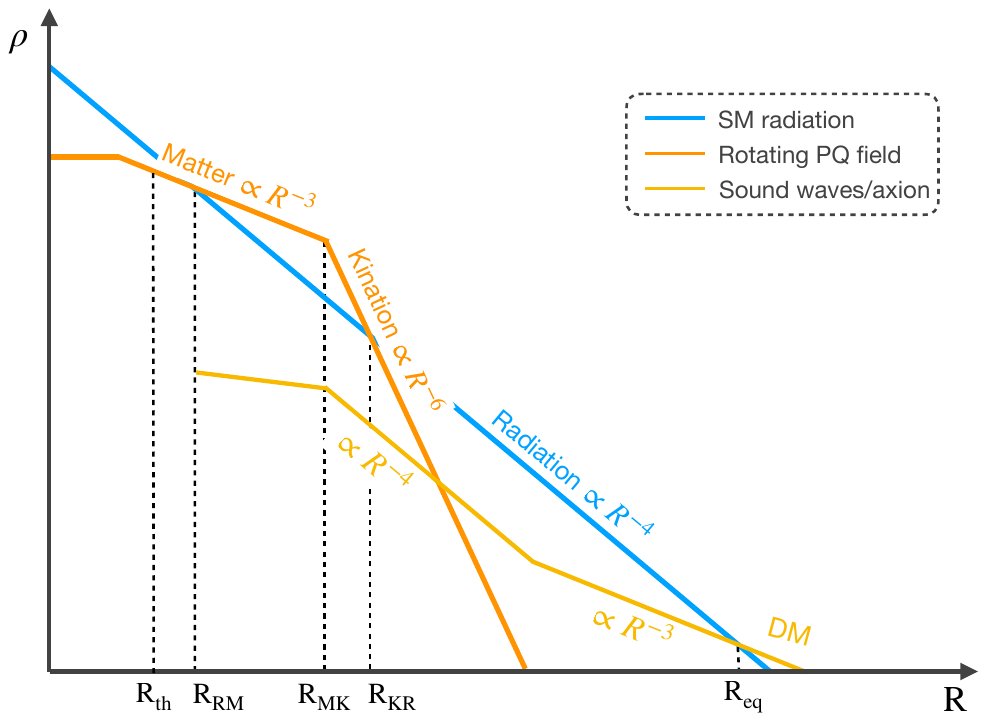}
	\caption{A schematic showing post-inflationary cosmology in nearly quadratic models.
    The energy density in the rotating axion (orange line) scales as matter initially and transitions to kination when the radius of rotation reaches the potential minimum.
    This may lead to a period of axion domination in the early universe. The sound-wave mode of the fluctuations (yellow line) around the rotating zero mode is produced from cosmic perturbations and evolves. 
    The sound waves become axion fluctuations at the bottom of the potential.       
    The energy density of the axion fluctuations
    initially dilutes like radiation,
    and becomes non-relativistic as the mass of the axion becomes important.
    The axion fluctuations then transition to matter-like behavior and can make up the entirety of dark matter today. There can be nonlinear evolution of the axion fluctuations around this transition as discussed in Sec.~\ref{sec:NL}.} 
\label{fig:cosmo_schematic}
\end{figure}

In both the log-potential model and the two-field model, the potential of $r$ is indeed nearly quadratic for $r \gg \ndw f_a$. Then the rotation behaves as matter, $\rho_\theta\propto R^{-3}$. After $r$ reaches the minimum, $\rho_\theta \propto R^{-6}$.
In \fig{fig:cosmo_schematic}, we show a schematic of the scaling of the radiation energy density and the rotation energy density for the nearly quadratic potential of $r$. Because of the initial matter scaling, the rotation can dominate the universe. We call the transition point from radiation domination to matter domination RM, from matter domination to kination domination MK, and from kination domination to radiation domination KR. Even when the rotation does not dominate the universe, we call the transition point of the energy scaling of the rotation from matter to kination MK.

\subsection{Thermalization constraints}

A larger initial field value $r_i$ leads to a larger yield $Y_{\theta}$ but also a smaller interaction rate between the PQ field and the thermal bath at a given temperature. We discuss in this subsection the constraint on the maximal values of $Y_\theta$ allowed by successful thermalization. These constraints will be used in Sec.~\ref{sec:implications} to derive the left boundary of the orange-shaded region in Fig.~\ref{fig:AMM_cogen_rP1}. The dashed green lines are the extensions of the allowed regions if there exist more efficient thermalization channels than those considered in this work.

In the case of the QCD axion, the PQ field can interact with the gluons via one-loop corrections, with the interaction rate given by~\cite{Bodeker:2006ij,Laine:2010cq,Mukaida:2012qn}
\begin{align}
\label{eq:Gamma_g}
    \Gamma_g = \frac{b \ndw^2 T^3}{r^2},
\end{align}
where $b \simeq 10^{-2} \alpha_3^2 \simeq 10^{-5}$. This interaction rate decreases slower (faster) than the Hubble scale when $r > \ndw f_a$ (after $r$ reaches $\ndw f_a $). Therefore, successful thermalization must occur before $r$ reaches $\ndw f_a$. This gives an upper bound on the yield
\begin{align}
\label{eq:Y_max_th}
 Y_\theta \lesssim 
 400 \, \ndw^{-1} 
 \left( \frac{b}{10^{-5}} \right)^3 
 \left( \frac{m_r}{\rm GeV} \right)
 \left( \frac{10^8 \GeV}{f_a} \right)^4
 \left( \frac{g_*}{g_{\rm MSSM}} \right)^{5/2} .
\end{align}

If the energy is dominated by the $P$ field, the thermalization of the radial mode reheats the universe and creates entropy. In this case, the maximum possible yield is achieved and is given by $Y_\theta = \rho_{\rm th} / (m_r s(T_{\rm th}) ) = 3 T_{\rm th} / 4 m_r$. Here, we assume that the rotational energy is comparable to that of the radial mode, i.e., an ellipticity of order unity. The thermalization temperature $T_{\rm th}$ is calculated by solving $\Gamma_g = 3H$ and $\rho_{\rm th}=m_r^2 r_{\rm th}^2 = \pi^2 g_* T_{\rm th}^4 / 30$. The resultant upper bound on the yield is 
\begin{align}
\label{eq:Y_max_th_dom}
 Y_\theta \lesssim 
 200 \, \ndw^{-1/3}
 \left( \frac{b}{10^{-5}} \right)^{1/3} 
 \left( \frac{\rm TeV}{m_r} \right)^{1/3}
 \left( \frac{g_{\rm MSSM}}{g_*} \right)^{1/2} .
\end{align}

More efficient thermalization can result from additional couplings. For example, if there exists a coupling between $P$ and heavy fermions, $y P \psi \bar\psi$, the thermalization rate is
\begin{align}
    \Gamma_\psi = b_\psi y^2 T ,
\end{align}
where $b_\psi \simeq 0.1$. At a given temperature, such a thermalization channel is possible only if the fermions $\psi \bar\psi$ are in the bath, i.e., $m_\psi = y r \lesssim T$, which leads to an upper bound on the thermalization rate
\begin{align}
    \Gamma_\psi \lesssim \frac{b_\psi T^3}{r^2} .
\end{align}
The maximum possible rate gives an upper bound on the allowed yield equivalent to that given by Eqs.~(\ref{eq:Y_max_th}) and (\ref{eq:Y_max_th_dom}) but with $b \simeq 10^{-5}$ replaced by $b_\psi \simeq 0.1$.

We will impose these upper bounds on $Y_\theta$ when we consider the dark matter abundance from the AMM in Sec.~\ref{sec:DM}.

%%%%%%%%%%%%%%%%%%%%%%%%%%%
\section{Fluctuations of Axion Rotation}
\label{sec:fluctuations}
%%%%%%%%%%%%%%%%%%%%%%%%%%%%

In this section, we discuss the evolution of the fluctuations of the rotating field. We show the equation of motion of the field perturbations as well as the perturbations of fluids using the formulation developed in~\cite{Co:2021lkc,Eroncel:2022vjg,Harigaya:2023mhl,Eroncel:2024rpe,Co:2024oek,Eroncel:2025bcb}.

\subsection{Field and fluid  perturbations}

To compute the evolution of perturbations, we use the conformal time $\eta$ with the conformal Newtonian gauge,
\begin{align}
\label{eq:metricpert}
ds^2 = \scale^2\left( - \left(1+2\Psi\right)d\eta^2 + \left(1 + 2 \Phi\right) \delta_{ij}dx^idx^j  \right). 
\end{align}
We decompose the PQ field into the zero mode $(r_0,\theta_0)$ and fluctuations $(\delta r, \delta\theta)$,
\begin{align}
P = \frac{r}{\sqrt{2}} e^{i\theta} = \frac{r_0 + \delta r}{\sqrt{2}}  e^{i\left(\theta_0 + \delta \theta\right)}.
\end{align}
The equations of motion for the zero mode and fluctuations generically contain two modes that couple with each other. However, after the thermalization of the rotation and in the limit where the mass of the radial direction is much larger than the Hubble expansion rate, we may integrate out the heavier mode~\cite{Co:2021lkc}, which is not associated with the conserved charge and is dissipated. We obtain the following relations~\cite{Co:2024oek},%
\footnote{For the two-field model, $P$ does not have a canonical kinetic term. $r$ used here is the field after the change of the variable such that the kinetic term of $\theta$ is $\dot{\theta} r^2/2$. See~\cite{Co:2024oek} for the expression for the potential in the two-field model, where $r$ here is called $F$. See~\cite{Harigaya:2023mhl} for the expressions for non-canonical kinetic terms.}
\begin{align}
\label{eq:relation}
\frac{1}{\scale^2}{\theta_0'}^2 & = \frac{V_r(r_0)}{r_0},\\
\label{eq:f_def}
\frac{\delta r}{r_0} & = g(r_0) \times  \left( \frac{\delta \theta'}{\theta_0'}-\Psi\right),~~ 
g(r_0) \equiv  \frac{2 V_r(r_0)/r_0}{V_{rr}(r_0)-V_r(r_0)/r_0}.
\end{align}
Here prime denotes derivative with respect to $\eta$ and the subscript $r$ denotes that with respect to $r$.
The remaining mode corresponds to the charge density and a phonon mode, i.e., sound waves in the superfluid; see Sec.~\ref{sec:phonons}.
The equation of motions of them take the form of the equation for current conservation,
\begin{align}
\label{eq:n0eq}
 n' & = -3 {\cal H} n, \\
\label{eq:n1eq}
{\delta}n' &=  - 3 {\cal H} \delta n + \partial_i j_i,   
\end{align}
where
\begin{align}
n = & \frac{1}{\scale} \theta_0' r_0^2,\\
\delta n = & n \left( 2 \frac{\delta r}{r_0} + \frac{\delta \theta'}{\theta_0'} -\Psi + 3\Phi  \right),\\
j_i = & \frac{1}{\scale} r_0^2 \partial_i \delta \theta.
\end{align}
For the two-field model and the log-potential model, 
the potential is nearly quadratic for $r_0 \gg \ndw f_a$, and hence, $g(r_0) \gg 1$, indicating that the fluctuation is dominantly in $\delta r/r_0$.
The fluctuations of the rotation are expected to behave as those of a matter fluid at first order in perturbations. 
On the other hand, around the bottom of the potential, $g(r_0 \approx \ndw f_a) \ll 1$, which means that the fluctuation in $\delta r/r_0$ is negligible.
The fluctuations of the rotation are expected to behave as those of a kination fluid at first order in perturbations.

We can confirm the intuition above by obtaining fluid equations from field equations. In particular, we compute the energy density $\rho$, the pressure $p$, and the divergence of the fluid velocity $\Theta$ as a function of $r$ and $\theta'$ and use the equations of motions of the fields in Eqs.~\eqref{eq:n0eq} and \eqref{eq:n1eq} along with the relations in Eqs.~\eqref{eq:relation} and \eqref{eq:f_def} to find
\begin{equation}
    w \equiv \frac{p}{\rho} = \frac{r_0 V_r-2V}{r_0 V_r + 2V} 
\end{equation}
for the zero mode and
\begin{align}
\delta' &= \frac{w'}{1+w} \delta- 3(1+w)\Phi'-(1+w) \Theta, \\
\Theta' &= - {\cal H} \left(1 - 3 c^{2}_{s}\right) \Theta - \partial_i^2 \Psi - \frac{c^{2}_{s}}{1+w} \partial_i^2\delta, \\
w' &= 3{\cal H}(1+w) \left(w - \frac{1}{1+2g} \right),\\
\label{eq:cs}
c_s^2 &=  w -  \frac{w'}{3 {\cal H} (1+w )} = \frac{1}{1+2g}
\end{align}
for the perturbations~\cite{Co:2024oek}, where $\delta = \delta \rho / \rho$. To compute $w$, a constant term should be added to $V$ so that it nearly vanishes at the minimum of the potential. The perturbation equations are nothing but those of an adiabatic fluid with an equation of state parameter $w$, its time derivative $w'$, with the adiabatic speed of sound $c_s$ given by a combination of them.
Note that the fluid equations are obtained without averaging over cycles of periodic motion, unlike the case for oscillating scalar fields.

\begin{figure}
    \centering
    \includegraphics[width=\linewidth]{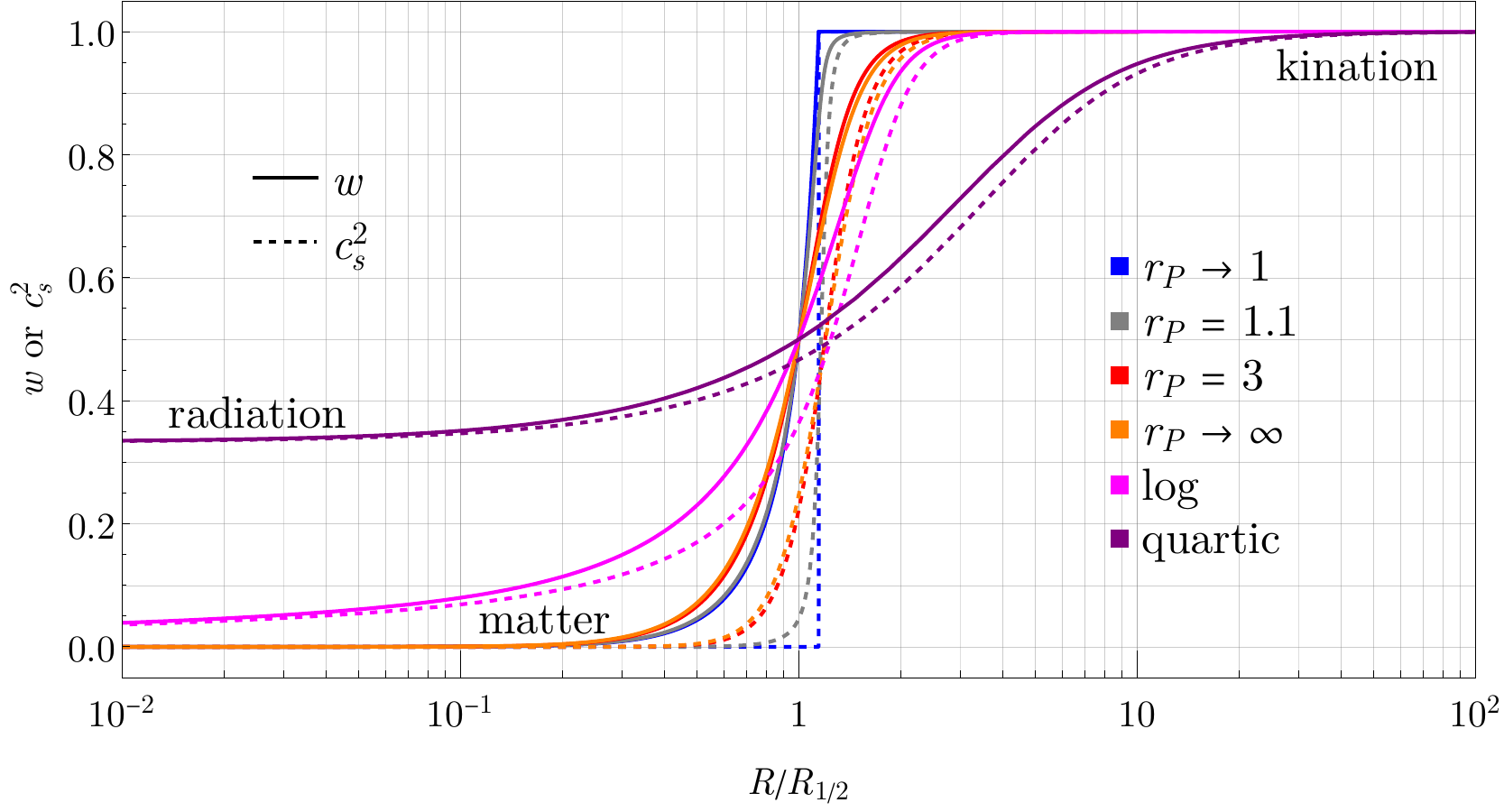}
    \caption{The equations of states for the two-field model with various $r_P$ values as labeled, for the log potential, and for the quartic potential. The scale factor $\scale_{1/2}$ is defined when the equation of state is $w=1/2$.}
    \label{fig:w_csSq}
\end{figure}

In Fig.~\ref{fig:w_csSq}, we show $w$ and $c_s$ as a function of $R$ for the log-potential model, the two-field model with different values of $r_P$, and the quartic model. We define $R_{1/2}$ as the scale factor when $w=1/2$. For $R\ll R_{1/2}$ where $r_0 \gg \ndw f_a$, the log-potential and two-field models have $w\sim c_s^2\sim 0$, and the fluctuations indeed behave as those of matter. The possible growth of fluctuations during the matter-like phase may affect the production of axion dark matter through the AMM.
The quartic model has $w\simeq c_s^2\simeq 1/3$ for $R\ll R_{1/2}$ and the perturbations behave as those of radiation. In all models, $w\simeq c_s^2 \simeq 1$ for $R\gg R_{1/2}$ and the perturbations behave as those of kination. 

\subsection{Perturbations as superfluid sound waves}
\label{sec:phonons}

\begin{figure}
    \centering
    \includegraphics[width=0.495\linewidth]{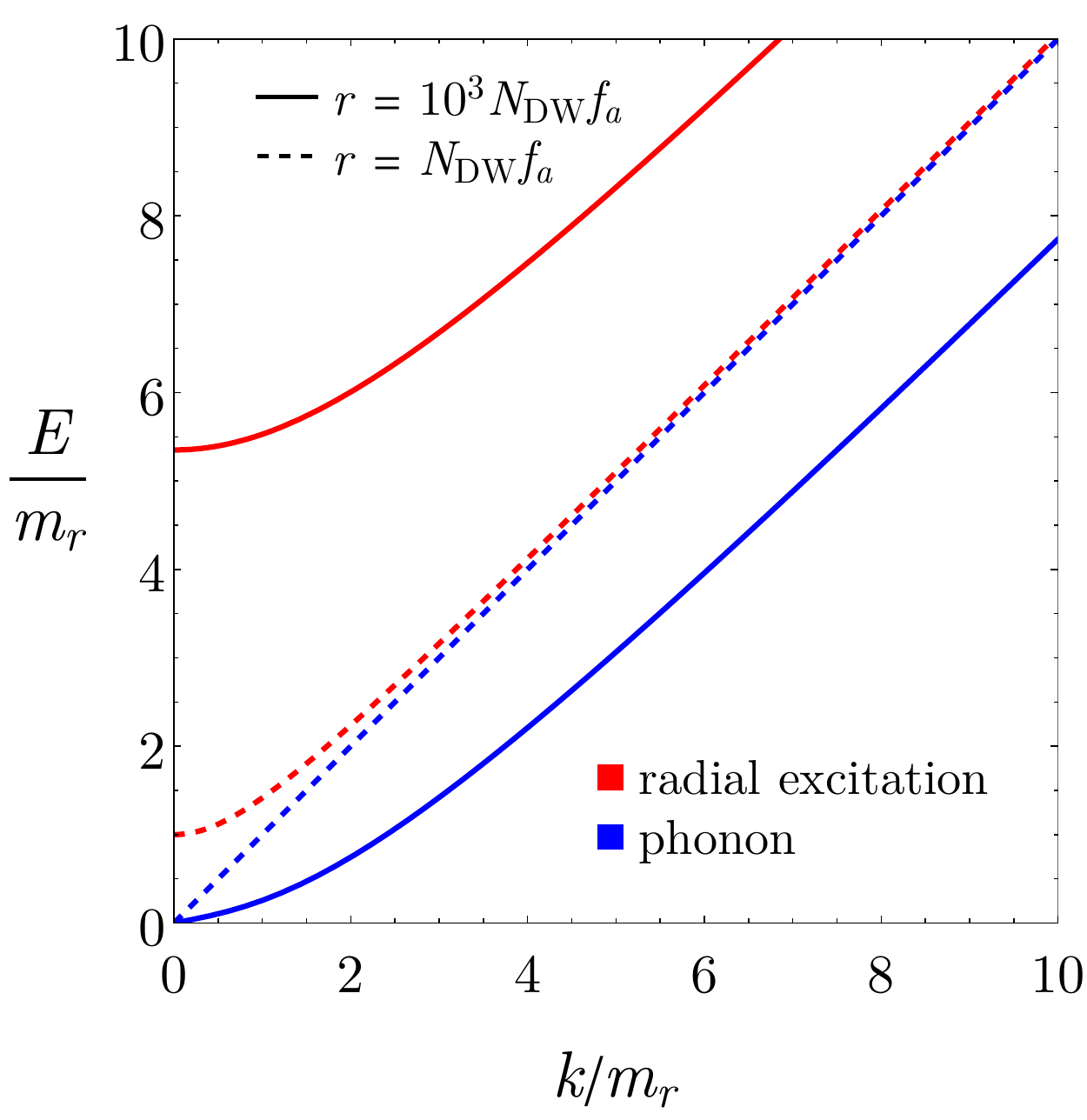}
    \caption{The dispersion relations of the two perturbation modes around the rotating background for the log-potential model. The solid lines correspond to $r= 10^{3}N_{\rm DW}f_a$ (nonzero $U(1)$ charges), while the dashed lines are for $r= N_{\rm DW}f_a$ (no $U(1)$ charges). In both cases, there is a gapless mode corresponding to phonons shown in blue, and the massive radial excitation mode is shown in red.}
    \label{fig:dispersion}
\end{figure}

The nature of the fluctuations in the axion rotation can be understood by spontaneous symmetry breaking involving the time-translational symmetry.
The rotation background with $\dot{\theta} = \omega \neq 0$ spontaneously breaks the $U(1)$ symmetry and the time translational symmetry into a diagonal $U(1)$ subgroup,
\begin{align}
\theta \rightarrow \theta + \alpha, ~~t \rightarrow t - \frac{\alpha}{\omega},~~\alpha \sim \alpha + 2\pi.
\end{align}
The perturbations around the rotation background can be understood as the NGB associated with this symmetry breaking, which is a sound wave of the PQ charge density fluctuations. In Appendix~\ref{app:phonon}, we present a derivation of the dispersion relation including both the phonon and gapped modes, which is shown in Fig.~\ref{fig:dispersion} for the log-potential model.

At low momenta, the dispersion relation of the phonon obeys $E = c_s |{\bf k}|$, where
\begin{equation}
\label{eq:cs2_phonon}
c_s^2 = \frac{V_{rr}-V_{r}/r}{V_{rr}+3V_r/r} = \frac{1}{1+2g(r)}. 
\end{equation} 
Because of the involvement of the time-translational symmetry in the symmetry breaking, the dispersion relation deviates from that of a relativistic NGB and hence $c_s < 1$. When the rotation occurs at the body of the potential, $|\dot{\theta}| = \sqrt{V_r/r}$ is as large as the typical mass scale of the theory $m_r \sim V^{1/2}_{rr}$, and the deviation of $c_s$ from 1 should be significant, which can be confirmed from Eq.~\eqref{eq:cs2_phonon} and Fig.~\ref{fig:w_csSq}. For the two-field model, $c_s$ is even close to 0. 
When the rotation occurs at the bottom, on the other hand, $|\dot{\theta}| \ll m_r$ and we expect that the involvement of time-translational symmetry is not important and $c_s \simeq 1$, which can be also confirmed from Eq.~\eqref{eq:cs2_phonon} and Fig.~\ref{fig:w_csSq}.

From the sound-wave picture, it is clear that the interaction of the fluctuations with the thermal bath is suppressed by the smallness of their momentum comparable to the cosmological scale, and the fluctuations will not be affected by the interaction with the thermal bath at any stage in the early universe, including the pre-kination phase.

In order for stable sound-wave modes to exist, $c_s^2 \geq 0$, i.e., $V_{rr}-V_r/r \geq0$ is necessary. This means that $P$ should have repulsive self interaction and the BEC of the PQ charges is a superfluid. 
We note that $c_s^2$ may become negative if the non-quadratic part of  the thermal potential dominates over that of the zero-temperature potential. We assume that $c_s^2$ is determined by the zero-temperature potential at RM and afterward, and is positive. If $c_s^2 <0$, fluctuations are exponentially enhanced by tachyonic instability and Q-balls can be formed~\cite{Coleman:1985ki,Kasuya:2010vq}. Although the Q-balls eventually decay~\cite{Chiba:2010ff,Co:2022qpr} as the zero-temperature potential does not admit Q-ball solutions, the enhanced fluctuations as well as the decay of the Q-balls can produce axion dark matter~\cite{Co:2019jts,Co:2022qpr}. The exponentially enhanced fluctuations may also cause the domain wall problem for the log-potential model with $\ndw >1$~\cite{Barnes:2022ren}, although whether or not domain walls actually form needs to be checked via lattice simulations.

\subsection{Perturbations as axion dark matter}
\label{sec:axion radiation}

During the kination phase, where $r_0$ is at the bottom of the potential, $\delta r/r_0$ is negligible, and the equation of motion of $\delta \theta$ is given by
\begin{align}
    \delta \theta '' + 2 {\cal H}\delta \theta' - \partial_i^2 \delta \theta + \theta' (- \Psi' + 3\Phi') = 0.
\end{align}
Because of the rapid decrease of $\theta'$ and the decrease of the metric perturbations far inside the horizon, the fourth term becomes negligible, and the comoving wavenumber $k \gg {\cal H}$, so $\delta \theta$ follows the equation motion of a free massless scalar in an expanding universe. Therefore, the second-order perturbations of the kination fluid, which contain the kinetic and gradient terms of $\delta \theta$, behave as radiation. The axion radiation energy density $\propto R^{-4}$ eventually exceeds the kination energy density $\propto R^{-6}$, after which the axions should not be considered as the fluctuations of the kination fluid. Since $\delta \theta$ simply follows the equation of motion of a free massless scalar, we may continue to use the radiation scaling of the fluctuations, as elucidated in~\cite{Eroncel:2025bcb}.

The fluctuation of the axion field 
$\delta \dot{\theta_a} = \ndw \delta \dot{\theta}$ is given by
\begin{equation}
\label{eq:delta relation}
    \delta \dot{\theta}_a = \frac{1}{2}\dot{\theta}_a \delta
\end{equation} 
during the kination phase.
The energy and number densities of the axion radiation are
\begin{align}
    \rho_a = & f_a^2 (\dot{\delta\theta_a})^2 = \frac{1}{4}\dot{\theta}_a^2 f_a ^2 \int \frac{{\rm d}k}{k}{\cal P}_\delta (k),\\
    n_a = & \frac{1}{4}\dot{\theta}_a^2 f_a ^2 \int \frac{{\rm d}k}{k }\frac{1}{k/\scale}{\cal P}_\delta (k) \equiv \int \frac{{\rm d}k}{k} \frac{{\rm d} n_a}{{\rm dln} k}(k),
\end{align}
respectively, where $k/\scale$ is the physical momentum. Note that ${\cal P}_\delta \propto R^{2}$ during the kination phase and hence $\rho_a \propto R^{-4}$ and $n_a\propto R^{-3}$.
Once the axion mass becomes larger than the physical wavenumber of the fluctuations, the axions become non-relativistic and behave as dark matter, as shown by the yellow line in Fig.~\ref{fig:cosmo_schematic}.
The possibility of axion dark matter from cosmic perturbations is noted in~\cite{Eroncel:2024rpe,Eroncel:2025bcb}, which discuss the evolution of axion field perturbations during the kination phase, although the dark matter abundance is not estimated. To estimate the resultant dark matter abundance, it is crucial to compute the evolution of $\delta$ before the kination phase, which is discussed in Sec.~\ref{sec:pert eq}.

\subsection{Evolution of perturbations}
\label{sec:pert eq}

We assume a tightly coupled radiation bath, which is justified when $T \gg $ MeV, and work in the regime where the second-order perturbations of the axion field are negligible.
Then the anisotropic stress tensor may be neglected and  $\Psi = - \Phi$. The evolution equations to be solved are given by 
\begin{align}
 \delta' = & \frac{\omega'}{1+\omega} \delta- 3(1+\omega)\Phi'-(1+\omega) \Theta, \\
\Theta' = & - {\cal H} \left(1 - 3 c^{2}_{s}\right) \Theta + \partial_i^2 \Phi - \frac{c^{2}_{s}}{1+\omega} \partial_i^2\delta,\\
 \delta_\gamma' = & - 4\Phi'- \frac{4}{3} \Theta_\gamma ,\\
\Theta_\gamma' = &  ~\partial_i^2 \Phi - \frac{1}{4} \partial_i^2\delta_\gamma,\\
  3\mathcal{H}\left(\Phi' + \mathcal{H} \Phi\right) = &~\partial_i^2 \Phi + 4 \pi G \scale^{2} \left(\rho \delta  + \rho_{\gamma} \delta_{\gamma} \right),
\end{align}
where $\rho_\gamma$,  $\delta_\gamma$, and $\Theta_\gamma$ are the energy density, its fluctuations, and the divergence of the fluid velocity of the radiation, respectively. Similarly,
$\rho$, $\delta$, and $\Theta$ are those of the rotation.
For our computation, it is more convenient to use the scale factor $\scale$ as a time variable. Going to the Fourier space, we find 
\begin{align}
 \frac{\partial\delta}{\partial \scale} = & \frac{\partial\omega/\partial \scale}{1+\omega} \delta- 3(1+\omega) \frac{\partial\Phi}{\partial \scale}-\frac{1+\omega}{\scale^2 H} \Theta , \\
\frac{\partial\Theta}{\partial \scale} = & - \frac{1}{\scale} \left(1 - 3 c^{2}_{s}\right) \Theta - \frac{k^2}{ \scale^2 H} \Phi + \frac{c^{2}_{s}}{(1+\omega)\scale^2H} k^2\delta,\\
 \frac{\partial\delta_\gamma}{\partial \scale} = & - 4 \frac{\partial\Phi}{\partial \scale}-\frac{4}{3 \scale^2 H} \Theta_\gamma ,\\
\frac{\partial\Theta_\gamma}{\partial \scale} = &   - \frac{k^2}{\scale^2 H} \Phi + \frac{k^2}{4\scale^2 H} \delta_\gamma,\\
  3 \scale \frac{\partial \Phi}{\partial \scale} = & - \Phi \left(\left(\frac{k}{\scale H}\right)^{2} +3\right) +  \frac{3}{2}\left(\frac{ \rho \delta}{\rho + \rho_{\gamma}} + \frac{\rho_{\gamma} \delta_{\gamma}}{\rho + \rho_{\gamma}} \right).
\end{align}

\begin{figure}
    \centering   \includegraphics[width=0.9\linewidth]{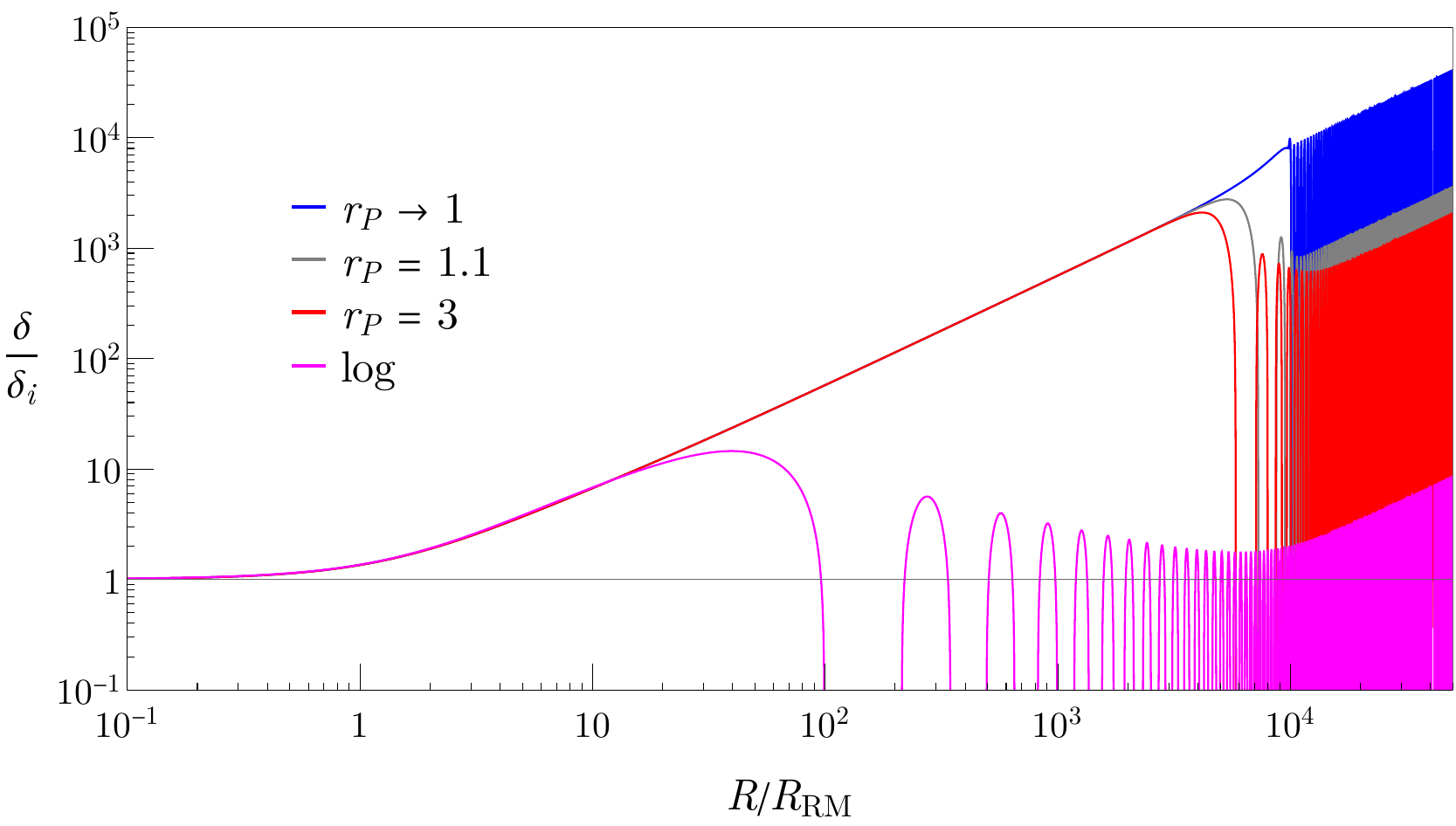}
    \caption{The evolution of the density perturbations of the axion rotation for the mode that enters the horizon at the transition from the radiation to matter domination (RM). We take $\scale_{\rm KR}/\scale_{\rm RM}= 10^6$. The pink line shows the evolution for the log-potential, and the blue, gray, and red lines show that of the two-field model with $r_P=1.001$, $1.1$, and $3$, respectively. $r_P= 1.001$ exhibits the maximal possible growth $\scale_{\rm MK}/\scale_{\rm RM}\simeq 10^4$ during the matter phase, while the growth for other cases is less.}
    \label{fig:delta_evolution}
\end{figure}

For the log-potential model, $w$ and $c_s$ are given by
\begin{align}\label{eq:cs_logPotential}
    w = \frac{\left(\frac{r}{\ndw f_a}\right)^2-1}{4 \left(\frac{r}{\ndw f_a}\right)^2{\rm ln}\left(\frac{r}{\ndw f_a}\right) +1 - \left(\frac{r}{\ndw f_a}\right)^2 } ,~~c_s^2 =  \frac{1}{1+4 \ln \paren{\frac{r}{\ndw f_a}}}.
\end{align}
For a large range of values of $r \gg \ndw f_a$, $ w^{1/2} \simeq c_s   \sim  0.1$ with a slow evolution because of the logarithmic dependence on $r$.
Given the smallness of $w$ in the majority of the pre-kination phase, the background evolution is fairly close to matter.  
However, an important consequence of the nonzero $c_s$ is a nonzero Jeans scale, which stops the growth of perturbations before the end of the matter-like phase. 
The scale factor at which this occurs, denoted by $\scale_s(k)$, can be estimated using
\begin{align}
    k^2 = \frac{\scale_s^2 H_s^2}{c_s^2(\scale_s)} \rightarrow c_s^2 (\scale_s) \approx 
    \frac{\scale_k}{\scale_s (k)},
\end{align}
where $\scale_k$ is the scale factor when the mode $k$ enters the horizon.
For $\scale> \scale_s (k)$, the mode starts oscillating.
From the Poisson equation deep inside the horizon, we obtain \cite{Mukhanov:2005sc}
\begin{align}
    \delta \approx \paren{\frac{k}{\scale H}}^2 \Phi,~~\Phi  \propto \frac{1}{(c_s k \eta)^{5/2}} J_{5/2} (c_s k \eta) \xrightarrow{c_s k \eta \gg 1} \frac{1}{(c_s k \eta)^{3}} \sin(c_s k \eta) .
\end{align}
Therefore, 
\begin{align}
    \delta \propto \frac{1}{c_s^3 \scale^{1/2}} \propto \frac{1}{\scale^{1/2}}.
\end{align}
Here we have used the fact that $c_s$ is a slowly changing function of $\scale$ for most of the matter-like phase.%
\footnote{More precisely, $\delta \propto R^{-(1-3w)/2}$. Before the MK transition, $w \sim 0.1$, which would result in a reduced suppression in $\delta$ compared to our estimate using $w =0$. The reduced suppression, however, does not change our estimation of the axion dark matter abundance. 
}
Then we can estimate the overall growth of the fluctuations during the matter-like phase to be
\begin{align}
    \frac{\delta(k,\scale_{\rm MK})}{\delta_{i}(k)} \approx \frac{\scale_s (k)}{\scale_k} \frac{\scale_s^{1/2} (k)}{\scale_{\rm MK}^{1/2}} \xrightarrow{k=k_{\rm RM}} \paren{\frac{\scale_{\rm RM}}{\scale_{\rm MK}}}^{1/2} \frac{1}{c_s^3 (\scale_s (k_{\rm RM}))}.
\end{align}
For example, for $\scale_{\rm KR}/\scale_{\rm RM} = 10^6$, using Eq.~\eqref{eq:cs_logPotential} we get $c_s^{-2} (\scale_s (k_{\rm RM})) \approx 35$, giving an enhancement in the mode re-entering at ${\rm RM}$ to be $ \frac{\delta(k_{\rm RM},\scale_{\rm MK})}{\delta_{i}(k_{\rm RM})}  \approx 2$. 
This is consistent with the numerical results that we show in \fig{fig:delta_evolution}.
In conclusion, the growth of perturbations during the matter-like phase is fairly modest in the log-potential model.

In the two-field model, on the other hand, $c_s \ll 1$, and we expect efficient growth of the perturbations inside the horizon during the matter phase if the rotation dominates.
However, finite $c_s$ for $r_P > 1$ can still suppress the growth to some extent.
The quartic model has $c_s=1/3$ before the kination phase and there is no growth of $\delta$.

We solve the evolution equations numerically
taking the adiabatic initial condition where $\delta_i = 3\Phi_i/2$ and $\delta_{\gamma_i} = \Phi_i$. The qualitative results are the same for isocurvature modes.
In \fig{fig:delta_evolution}, we show the evolution of $\delta$ for the mode that enters the horizon at RM for $\scale_{\rm KR}/\scale_{\rm RM} = 10^{6}$, for which $\scale_{\rm MK}/\scale_{\rm RM}\simeq 10^4$. 
In the log-potential model, the perturbation exhibits growth after entering the horizon, but soon after, it begins to oscillate and is suppressed due to $c_s \sim 0.1$. The scaling of $\delta$ inside the sound horizon and the overall growth is consistent with the analytical estimate given above.  
The two-field model exhibits more efficient growth, but unless $r_{P}-1 < 10^{-3}$ the perturbation begins to oscillate rapidly and is suppressed before MK.%
\footnote{
The rapid oscillations after the growth produce gravitational waves~\cite{Inomata:2019ivs,Harigaya:2023mhl}.}
For larger $\scale_{\rm KR}/\scale_{\rm RM}$, even $r_P=1.001$ does not exhibit maximal growth for $k_{\rm RM}$. This is because
$\int {\rm d}\eta c_s k_{\rm RM}$
becomes larger than one just before MK, and the resultant rapid oscillations lead to suppression.
In all models, $\delta$ grows linearly during the kination phase, which can be seen for $\scale/\scale_{\rm RM}\gtrsim 10^4$ in \fig{fig:delta_evolution}. 
These results are used in Sec.~\ref{sec:DM} to compute the axion dark matter abundance.

\section{Axion dark matter from fluctuations}
\label{sec:DM}

In this section, we compute the dark matter abundance produced from the fluctuations of the axion rotation and discuss implications for the parameter space of axion models.
We focus on the case with nearly quadratic potentials, for which axion rotation may dominate the universe.
The axion abundance for the quartic model is the same as that for the quadratic case without axion domination (discussed in Sec.~\ref{sec:subdominant}), with MK interpreted as the moment when the scaling of the rotation energy transitions from  radiation-like  to kination-like.

The axion abundance can be estimated following the procedure laid out in Sec.~\ref{sec:axion radiation}.
The number density spectrum is given by
\begin{align}
\label{eq:naspectrum}
   \frac{{\rm d} n_a}{{\rm dln} k}(k)  =  \frac{1}{4}\dot{\theta}^2_a f_a^2\frac{1}{k/\scale} {\cal P}_\delta(k) = &  {\cal P}_{\delta_i}(k)  \frac{ \dot{\theta}^2_{a,\rm MK}f_a^2}{4H_{\rm MK}} \times \frac{{\cal P}_\delta(k)}{{\cal P}_{\delta_i}(k)}  \frac{H_{\rm MK}}{k/\scale_{\rm MK}} \left(\frac{\scale_{\rm MK}}{\scale }\right)^5 \\ \nonumber 
    =&  {\cal P}_{\delta_i}(k)  \frac{ \dot{\theta}^2_{a,\rm MK}f_a^2}{4H_{\rm MK}} \times F(k) \times \frac{\scale_{\rm MK}^3}{\scale^3},
\end{align}
where
${\cal P}_{\delta_i}$ is the initial power spectrum of $\delta$ far outside the horizon and
\begin{align}
\label{eq:F}
    F(k)\equiv & \frac{{\cal P}_{\delta}(k)_{R \gg R_{\rm MK}}}{{\cal P}_{\delta_i}(k)} \frac{R_{\rm MK}^2}{R^2} \frac{H_{\rm MK}}{k/\scale_{\rm MK}}.
\end{align}
Note that $F(k)$ becomes constant far inside the horizon during the kination phase since ${\cal P}_{\delta}(k)_{R \gg R_{\rm MK}} \propto \scale^2$.
The mode that enters the horizon at MK, where $k= \scale_{\rm MK} H_{\rm MK}\equiv k_{\rm MK}$,  has $F=1$, so the factor $F$ indicates the importance of the mode $k$ in comparison with the mode that enters the horizon at MK.

The modes that enter the horizon after MK have
\begin{equation}
 {\cal P}_{\delta}(k \ll k_{\rm MK})_{R \gg R_{\rm MK}} = {\cal P}_{\delta_i}(k) \times \left(\frac{\scale}{\scale_k} \right)^2.
\end{equation}
The number density spectrum is thus given by
\begin{equation}
\label{eq:na_postMK}
    \frac{{\rm d} n_a}{{\rm dln} k}(k \ll k_{\rm MK})  = {\cal P}_{\delta_i}(k)  \frac{ \dot{\theta}_{a,\rm MK}^2 f_a^2 }{4 H_{\rm MK}} \frac{H_{\rm MK}}{k/\scale_{\rm MK}} \frac{\scale_{\rm MK}^2}{\scale_k^2} \times \frac{\scale_{\rm MK}^3}{\scale^3}.
\end{equation}
The modes that enter the horizon before MK can experience non-trivial evolution before MK and hence $F$ needs to be in general computed numerically.

\begin{figure}
    \centering   \includegraphics[width=0.9\linewidth]{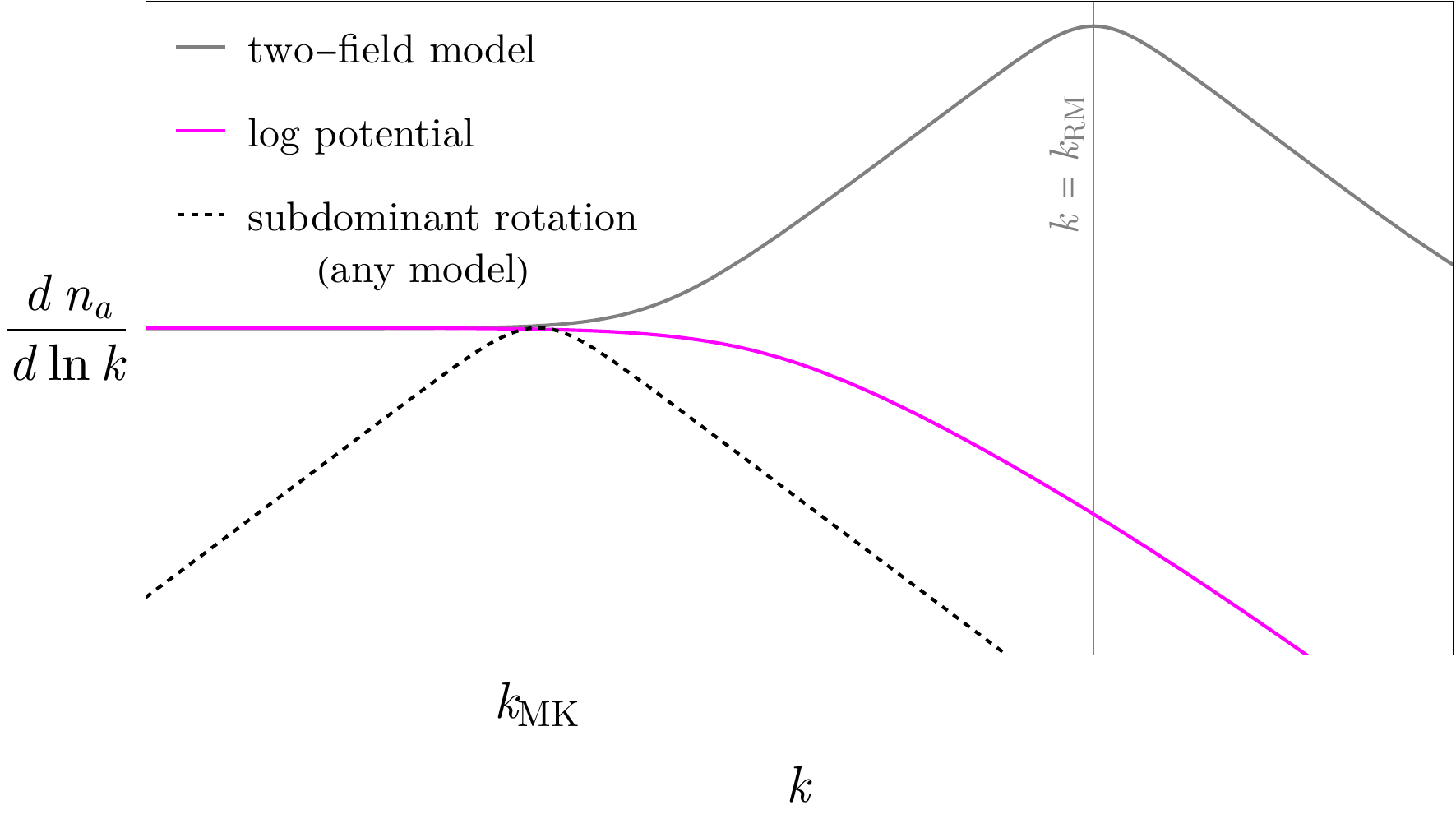}
    \caption{A schematic figure of the axion number density spectrum ${\rm d} n_a/{\rm dln} k$ as a function of wavenumber~$k$ in log-log scale. The gray (magenta) line is for the two-field model (log-potential model) in the case where the axion rotation dominates the energy density of the universe. The case of no domination by the axion rotation is shown by the black dashed line, which is applicable to all models.}
    \label{fig:dna_dlnk}
\end{figure}

In the following, we consider both cases where the axion rotation does not and does dominate the energy density of the universe.
We assume that the primordial perturbation is approximately flat for the modes that we are interested in and take ${\cal P}_{\delta_i}(k)= {\cal P}_{\delta_i}$.
A schematic picture summarizing the spectrum of the axion number is shown in Fig.~\ref{fig:dna_dlnk}.

\subsection{Subdominant axion rotation}
\label{sec:subdominant}

When the rotation is subdominant, the growth of the fluctuation during the matter phase is at the most logarithmic. Therefore, the contribution to the axion abundance from the modes that enter the horizon before MK in Eq.~\eqref{eq:naspectrum} is the largest at the smallest $k$, i.e., $k_{\rm MK}=\scale_{\rm MK}H_{\rm MK}$, and we obtain
\begin{equation}
\label{eq:na_sub}
 n_a = {\cal P}_{\delta_i} \frac{ \dot{\theta}^2_{a,{\rm MK}}f_a^2}{4 H_{\rm MK}}\times  \frac{\scale_{\rm MK}^3}{\scale^3}.
\end{equation}
The contribution to the axion abundance from the modes that enter the horizon after MK in Eq.~\eqref{eq:na_postMK} is the largest at the largest $k$, i.e., $k_{\rm MK}$, since $k\propto \scale_{k}^{-1}$ during radiation domination.  The resultant abundance is also given by Eq.~\eqref{eq:na_sub}.
The number density spectrum of the axion is shown by the black dashed line in Fig.~\ref{fig:dna_dlnk}.

\subsection{Axion rotation domination}

When the rotation dominates the universe, the axion abundance is model-dependent. Let us first discuss model-independent contributions.
The modes that enter the horizon during kination domination follow the scaling $k\propto R_k^{-2}$, so the axion number density in Eq.~\eqref{eq:na_postMK} is the same for any $k$ that enters the horizon during kination domination, which corresponds to the flat part of the gray and pink lines in Fig.~\ref{fig:dna_dlnk}. 
The axion abundance produced during the kination phase receives a mild log enhancement,
\begin{equation}
\label{eq:na_kin}
     n_{a,{\rm kin}}  =  {\cal P}_{\delta_i} \frac{ \dot{\theta}_{a,\rm MK}^2 f_a^2 }{4H_{\rm MK}} {\rm ln}\left(\frac{\scale_{\rm KR}}{\scale_{\rm MK}}\right)^2 \times \frac{\scale_{\rm MK}^3}{\scale^3}.
\end{equation}
After KR, $k \propto R_k^{-1}$, so the axion production from the modes that enter the horizon after KR is subdominant.

We next discuss model-dependent contributions.
$\delta$ can grow during the matter-dominated phase. The contribution to the number density of the axion from the modes that enter the horizon before MK
can be estimated by finding $k$ that maximizes $F(k)$ in Eq.~\eqref{eq:F}.

In the two-field model with $r_P\simeq 1$, $\delta$ grows linearly during the matter domination. Using $k = k_{\rm MK} (\scale_{\rm MK}/\scale_k)^{1/2}$ and the linear growth of $\delta$ after entering the horizon for $\scale_k > \scale_{\rm RM}$, we find that $F$ is maximized for the mode that enters the horizon at RM and is given by
\begin{align}
\label{eq:na_dom_rP1}
    n_{a,r_P\simeq 1} = &  {\cal P}_{\delta_i} \left(\frac{\scale_{\rm MK}}{\scale_{\rm RM}}\right)^{3/2} \frac{ \dot{\theta}^2_{a,\rm MK}f_a^2}{4 H_{\rm MK}} \times \frac{\scale_{\rm MK}^3}{\scale^3} \\ \nonumber
     = & {\cal P}_{\delta_i} \frac{\scale_{\rm KR}}{\scale_{\rm RM}} \frac{ \dot{\theta}^2_{a,\rm MK}f_a^2}{4 H_{\rm MK}} \times \frac{\scale_{\rm MK}^3}{\scale^3}.
\end{align}
The axion number density spectrum is illustrated by the gray line in Fig.~\ref{fig:dna_dlnk}.

\begin{figure}[t]
    \centering
    \includegraphics[width=0.9
    \linewidth]{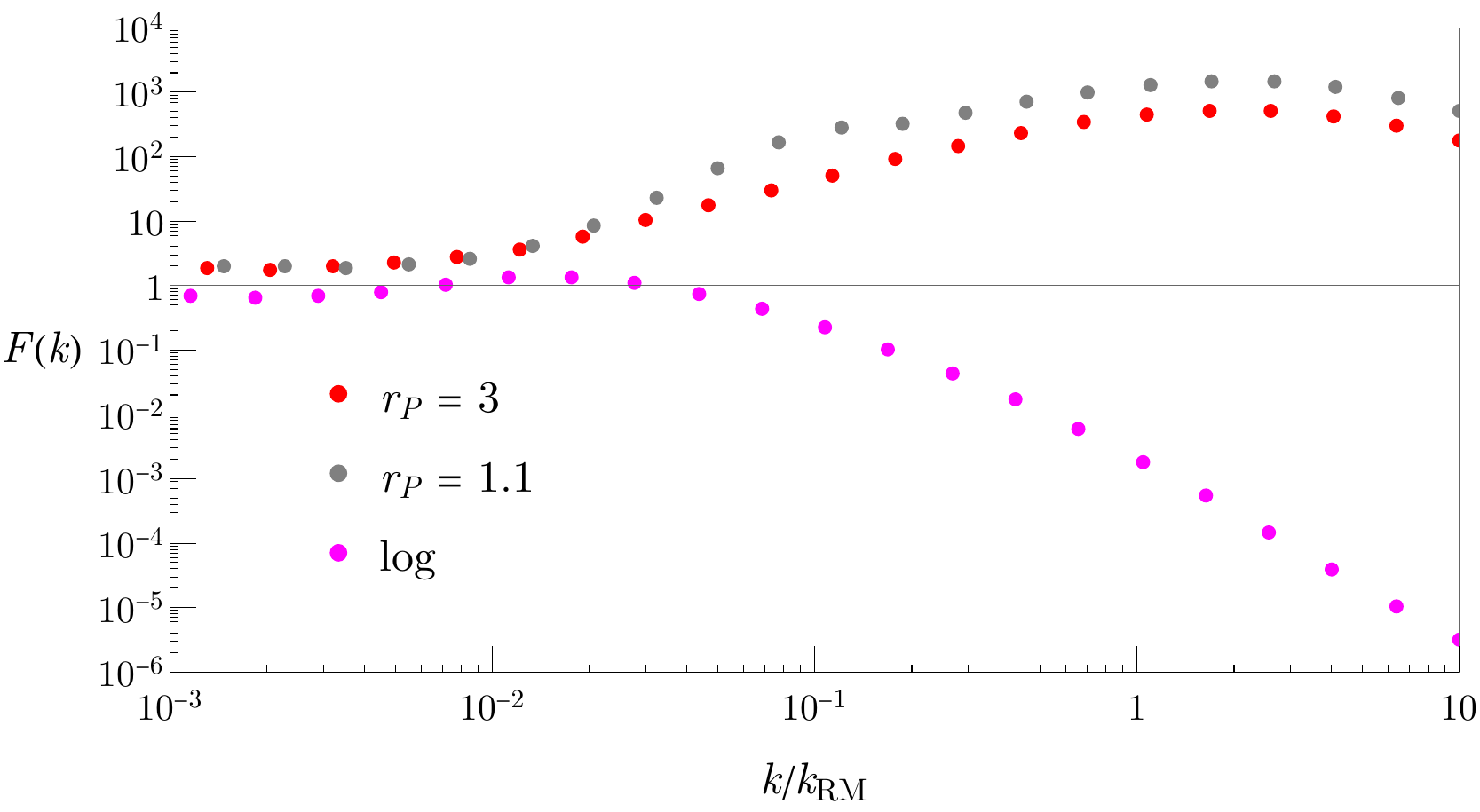}
    \caption{The factor $F(k)$ that indicates the importance of the mode $k$ in comparison with the mode that enters the horizon at MK. We take $R_{\rm KR}/R_{\rm RM}=10^6$, so $k_{\rm MK}/k_{\rm RM} \simeq 0.01$ and $\scale_{\rm MK}/\scale_{\rm RM}\simeq 10^4$. $k_{\rm RM}$ is the comoving wavenumber of the mode that enters the horizon at RM. The magenta dots show $F(k)$ for the log-potential model, while the gray (red) dots show it for the two-field model with $r_P = 1.1$ ($r_P = 3$).}
    \label{fig:F}
\end{figure}

For the log-potential model or the two-field model with $r_P > 1$, $F(k)$ needs to be computed numerically. In Fig.~\ref{fig:F}, we show $F(k)$ as a function of $k/k_{\rm RM}$ for $\scale_{\rm KR}/\scale_{\rm RM}=10^6$ for the log-potential model and the two-field model with $r_P=1.1$ or $3$.
The mode that enters the horizon at MK corresponds to $k/k_{\rm RM}\simeq 0.01$.
For the log-potential model, $F(k)$ at $k/k_{\rm RM} > 0.01$ is suppressed for the following reason. Although the modes that enter the horizon before MK can grow, because of the finiteness of $c_s\sim 0.1$, the perturbation stops growing shortly after entering the horizon and begins to be suppressed, as shown in Fig.~\ref{fig:delta_evolution}. The number density is then suppressed by the largeness of their wavenumbers from $F(k) \propto 1/k$. The flatness of $F(k)$ for $k/k_{\rm RM} < 0.01$ is consistent with the argument on the kination-dominated phase described above.
The axion number density for the log-potential model is therefore given by Eq.~\eqref{eq:na_kin}, with the spectrum given by the magenta line in Fig.~\ref{fig:dna_dlnk}.

\begin{figure}[t]
    \centering
    \includegraphics[width=0.9
    \linewidth]{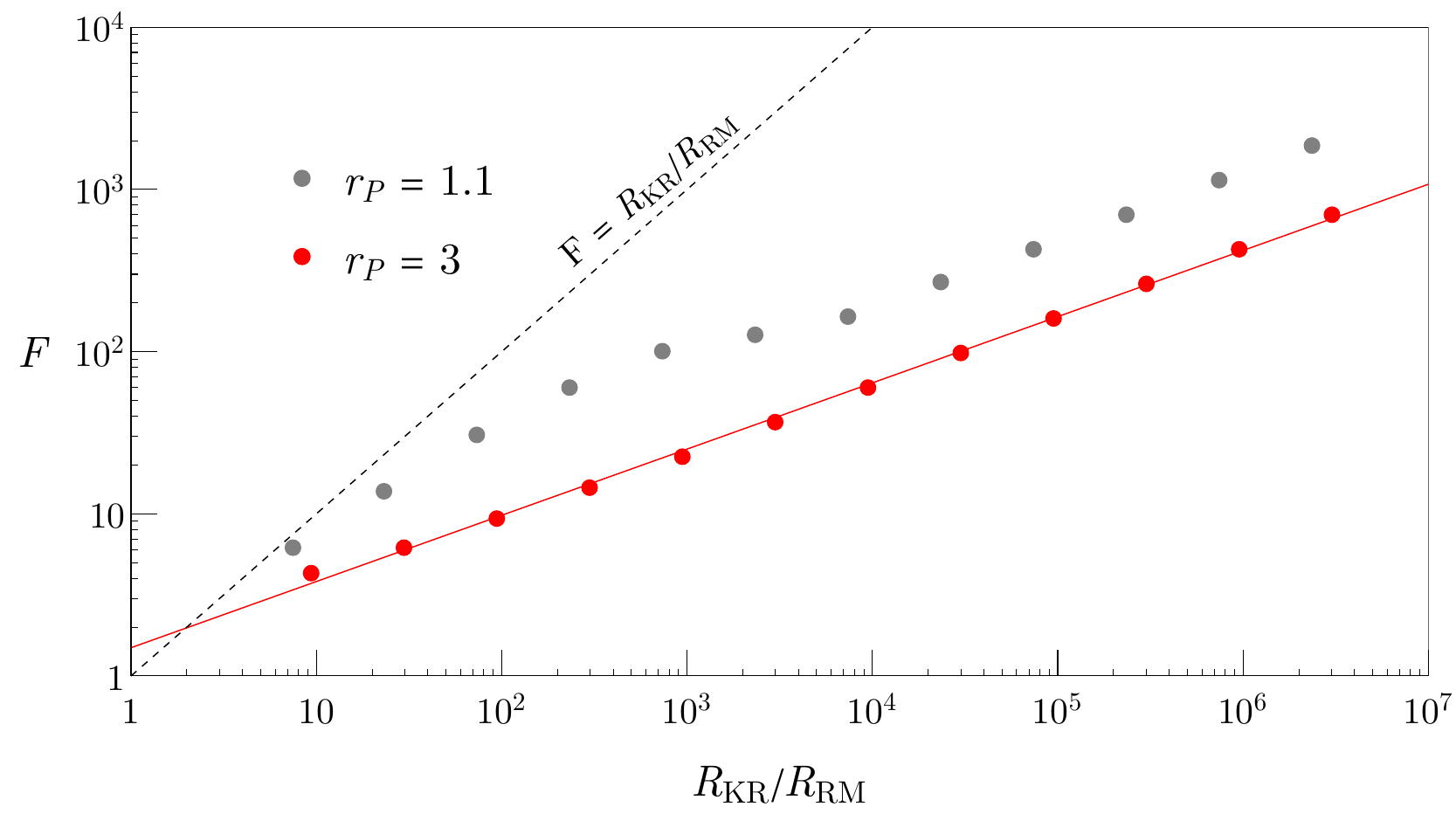}
    \caption{The factor $F$ for the mode that enters the horizon at RM as a function of $R_{\rm KR}/R_{\rm RM}$ for the two-field model with $r_P=1.1$ and $3$. The result for larger $r_P$ is similar to $r_P=3$.}
    \label{fig:Ftwo}
\end{figure}

$F$ of the two-field model is peaked at the mode that enters the horizon at RM because of the linear growth during the matter phase. However, because of the suppression just before MK, $F$ does not reach the maximal possible value $\scale_{\rm KR}/\scale_{\rm RM} = 10^6$ for $r_p >1$. 
In Fig.~\ref{fig:Ftwo}, we show $F(k_{\rm RM})$ as a function of $\scale_{\rm KR}/\scale_{\rm RM}$ for $r_P=1.1$ and $3$. $F(k_{\rm RM})$ is smaller than the maximal possible value shown by the dashed line, but is still much larger than 1, showing the importance of the matter phase. For larger $r_P$, we find that $F$ is smaller but only by a few ten percents. 
$F$ for $r_P >$ few can be fitted by
\begin{equation}
    F(k_{\rm RM}) \simeq 1.5 \times \left(\frac{\scale_{\rm KR}}{\scale_{\rm RM}}\right)^{0.41},
\end{equation}
which is shown by the red line in Fig.~\ref{fig:Ftwo}.
The corresponding axion number density is 
\begin{align}
\label{eq:na_dom_rPlarge}
    n_{a,r_P >{\rm few}} \simeq  1.5  \left(\frac{\scale_{\rm KR}}{\scale_{\rm RM}}\right)^{0.41} {\cal P}_{\delta_i}  \frac{ \dot{\theta}^2_{a,\rm MK}f_a^2}{4 H_{\rm MK}} \times \frac{\scale_{\rm MK}^3}{\scale^3}.
\end{align}
The spectrum follows the gray line in Fig.~\ref{fig:dna_dlnk}.

\subsection{Axion dark matter density}
\label{sec:DM density}

We now compute the axion abundance as a function of the model parameters.
The free parameters of the theory are $Y_\theta$, $f_a$, and $\ndw m_r$. We can trade the last one for $ d \equiv \left. \frac{\rho_\theta}{\rho_{\rm rad}} \right|_{\rm MK}$ using the relation
\begin{align}
    d \equiv \left. \frac{\rho_\theta}{\rho_{\rm rad}} \right|_{\rm MK} = \left( \frac{ \sqrt{128 \pi^2} g_*^{1/2}\ndw m_r Y_\theta^2}{ \sqrt{1215} f_a} \right)^{2/3} \simeq \left( \frac{ g_*^{1/2}\ndw m_r Y_\theta^2}{f_a} \right)^{2/3}. 
\end{align}
Key quantities to compute the axion abundance are given by
\begin{align}
    \rho_{\theta,{\rm MK}} = & \dot{\theta}_{\rm MK}^2 f_a^2 = \frac{1215 d^3 f_a^4}{128 \pi^2 g_* Y_\theta^4} \simeq 0.96 \frac{d^3 f_a^4}{g_* Y_\theta^4} , \\
    \rho_{{\rm rad,MK}} =& \frac{1215 d^2 f_a^4}{128 \pi^2 g_* Y_\theta^4} \simeq 0.96\frac{d^2 f_a^4}{g_* Y_\theta^4}, \\
    s_{\rm MK} = &\sqrt{\frac{1215}{128 \pi^2}} \frac{d^{3/2} f_a^3}{g_*^{1/2} Y_{\theta}^3} \simeq 0.98\frac{d^{3/2} f_a^3}{g_*^{1/2} Y_{\theta}^3}, \\
    H_{\rm MK} = &\sqrt{\frac{405}{128 \pi^2}} \frac{d (1+d)^{1/2} f_a^2}{ g_*^{1/2} Y_\theta^2 \mpl} \simeq 0.57 \frac{d (1+d)^{1/2} f_a^2}{ g_*^{1/2} Y_\theta^2 \mpl} , \\
    \frac{\scale_{\rm KR}}{\scale_{\rm MK}} = & d^{1/2} ,\\
    \frac{\scale_{\rm MK}}{\scale_{\rm RM}} \simeq & d,
\end{align}
where the last equation approximately holds for the two-field model or the log model with a long duration of rotation domination.

\subsubsection{Subdominant Rotation}
When the axion rotation does not dominate $(d <1)$, the number density of the axion is given by Eq.~\eqref{eq:na_sub}. Normalized by the entropy density $s$, we obtain
\begin{equation}
    Y_a \equiv \frac{n_{a}}{s} \simeq {\cal P}_{\delta_i} \frac{( \dot{\theta}_{a,{\rm MK}})^2 f_a^2}{ 4 s_{\rm MK} H_{\rm MK}} = {\cal P}_{\delta_i} \frac{\sqrt{3}\mpl}{4 f_a}  \sqrt{\frac{d}{1+d}} \times Y_\theta.
\end{equation}
The physical momentum of the axions normalized by the entropy density is
\begin{equation}
    y_k \equiv \frac{k/R}{s^{1/3}} \xrightarrow{k=k_{\rm MK}} \frac{H_{\rm MK}}{s_{\rm MK}^{1/3}} \simeq 0.57 \frac{\sqrt{d(1+d)} f_a}{g_*^{1/3} \mpl Y_\theta},
\end{equation}
where in the second part of the expression above, we have focused on $k_{\rm MK}$ mode that dominates the axion abundance.
The axion becomes non-relativistic when $k/\scale_{\rm NR} \sim m_a $. 
The corresponding temperature can be given as 
\begin{equation}
\label{eq:TNRsub}
    T_{\rm NR} = \left(\frac{45}{2\pi^2 g_*(T_{\rm NR})}\right)^{1/3}\frac{m_a}{y_k} = \frac{4}{\sqrt{3}} \frac{m_a \mpl Y_\theta}{ \sqrt{d(1+d)}f_a}  \left(\frac{g_*(T_{\rm MK})}{g_*(T_{\rm NR})}\right)^{1/3},
\end{equation}
which is required to be larger than about $5$ keV to satisfy the warmness constraint~\cite{Sitwell:2013fpa,Lopez-Honorez:2017csg}.
In setting the constraint it is assumed that the mass is constant at $T=T_{\rm NR}$, which may not be the case for the QCD axion whose mass depends on the temperature at $T > 100$ MeV. However, the constraint is relevant only if $T_{\rm NR} \ll 100$ MeV and we may safely assume a constant $m_a$ in setting the warmness constraint.

\subsubsection{Dominant Rotation}

When the axion rotation dominates, the axion abundance can be further enhanced. For the two-field model with $r_P\simeq 1$ and $r_P >$ few, the abundance is given by Eqs.~\eqref{eq:na_dom_rP1} and \eqref{eq:na_dom_rPlarge}, and we obtain
\begin{equation}
    Y_a  \simeq  {\cal P}_{\delta_i} \frac{\sqrt{3}\mpl}{4 f_a}  Y_\theta \times
    \begin{cases}
        d^{3/2} & r_P\simeq 1 \\
        d^{0.61}  & r_P > {\rm few}
    \end{cases} .
\end{equation}
The physical momentum of axions normalized by the entropy density is
\begin{equation}
    y_k \equiv \frac{k/R}{s^{1/3}} \xrightarrow{k=k_{\rm MK}} \frac{H_{\rm MK}}{s_{\rm MK}^{1/3}} d^{1/2} \simeq 0.57 \frac{d\sqrt{(1+d)} f_a}{g_*^{1/3} \mpl Y_\theta}.
\end{equation}
The axion becomes non-relativistic at a temperature
\begin{equation}
    T_{\rm NR} = \left(\frac{45}{2\pi^2 g_*(T_{\rm NR})}\right)^{1/3}\frac{m_a}{y_k} = \frac{4}{\sqrt{3}} \frac{m_a \mpl Y_\theta}{d \sqrt{1+d}f_a}  \left(\frac{g_*(T_{\rm MK})}{g_*(T_{\rm NR})}\right)^{1/3}.
\end{equation}

For the log model, the axion number density is given by Eq.~\eqref{eq:na_kin}, and we obtain
\begin{equation}
     Y_a  \simeq  {\cal P}_{\delta_i} \frac{\sqrt{3}\mpl}{4 f_a}  Y_\theta \times  \ln d.
\end{equation}
Since the production of the dark matter is dominated during the era of matter-kination transition, just as in the case where rotation is subdominant, the axion becomes non-relativistic at the same temperature as in Eq.~\eqref{eq:TNRsub}.

\subsubsection{Comparison with the KMM}

Let us compare the abundance with the KMM contribution~\cite{Co:2019jts,Co:2021rhi,Eroncel:2022vjg},
\begin{equation}
    Y_{a,{\rm KMM}} \simeq Y_\theta.
\end{equation}
The number density of the AMM contribution is larger than the KMM contribution if
\begin{equation}
\label{eq:fa_comparison}
    f_a \lesssim 2\times 10^{9}{\rm GeV} \times \frac{{\cal P}_{\delta_i}}{2\times 10^{-9}} \times 
    \begin{cases}(\ln d, d^{3/2},~d^{0.61}) \hspace{0.5 in} & \text{ dominant rotation} \\
    d^{1/2} \hspace{0.5 in} & \text{ subdominant rotation}
    \end{cases},
\end{equation}
where the $d$ dependence for the domination case is shown for the models: (log, two-field $r_P \rightarrow 1$, two-field $r_P \gg 1$). The result is universal for the subdominant case.
The range of $f_a$ in Eq.~\eqref{eq:fa_comparison} is consistent with the typical lower bound on the decay constant of the QCD axion, which is $f_a \gtrsim 4 \times 10^{8}$ GeV from the neutron star observations for the KSVZ model~\cite{Leinson:2021ety,Buschmann:2021juv} and $f_a \gtrsim 10^9$ GeV from the red giant observations for the DFSZ model~\cite{Capozzi:2020cbu,Straniero:2020iyi}.
This shows that the AMM can indeed be more important than KMM in a certain part of the parameter space of the QCD axion. 
This will be further illustrated in \fig{fig:QCDaxion}.

\subsection{Large fluctuations and nonlinear evolution}
\label{sec:NL}

In estimating the number density of axions so far, we implicitly assumed that $\delta \theta_a \lesssim 1$ when the physical momentum of the axions becomes smaller than the axion mass, so that the axion fluctuations simply oscillate around a single minimum of the potential after becoming non-relativistic. If $\delta \theta_a \gg 1$, the axion fluctuations spread over multiple periods of the potential and the estimation changes.
This occurs if
\begin{align}
 (\delta \theta_a^2)_{\rm NR}=  \frac{Y_a s_{\rm NR}}{m_a(T_{\rm NR}) f_a^2} = \frac{Y_a m_a(T_{\rm NR})^2}{y_k^3 f_a^2} \gg 1,
\end{align}
where in the first equality we used $n_a \simeq \delta\theta_a^2 f_a^2 k/R = \delta\theta_a^2 f_a^2 m_a $ at $T=T_{\rm NR}$.

After the momentum becomes smaller than $m_a$, initially $\delta \theta_a \, (k /R)> m_a$, and the dynamics of the axion fluctuations is determined by the kinetic and gradient terms. Therefore, the axion fluctuations continue to behave as radiation even though $k/R < m_a$, and $\delta \theta_a \, (k /R)$ decreases in proportion to $R^{-2}$.
Below a temperature $T_*$ defined by
\begin{equation}
   m_a^2(T_*) = (\delta \theta_a)^2 \frac{k^2}{R^2}= Y_a y_k \frac{s^{4/3}_*}{f_a^2}, 
\end{equation}
the dynamics of the fluctuations is governed by the potential energy. Since $\delta \theta_a \gg 1$, the axion fluctuations spread over multiple periods of the potential; small, closed, and boundary-less domain walls can be produced, but they decay immediately. Because of the nonlinear dynamics the axion number density will not be conserved around $T_*$, but for $T<T_*$ the number density is conserved again. (This dynamics is similar to what is described in Appendix E of~\cite{Gorghetto:2020qws}.)  The energy density of the axion fluctuations at $T=T_*$ is about $m_a^2 f_a^2$ and the energy per quantum is $m_a$, so the number density of axions is given by
\begin{align}
    Y_{a,\delta \theta_{a,{\rm NR}}\gg 1} = \frac{m_a(T_*)f_a^2}{s_*} = & (Y_a y_k)^{3/4} \frac{f^{1/2}_a}{m_a^{1/2}(T_*)} \nonumber \\
    =& Y_a \times \left(\frac{m_a(T_{\rm NR})}{\delta \theta_{a,{\rm NR}} m_a(T_*)} \right)^{1/2},
\end{align}
where $Y_a$ on the right-hand side matches the estimate given in Sec.~\ref{sec:DM density}.
The reduction in the number density can be attributed to the nonlinear evolution around $T_*$ that may violate the number density conservation.

The warmness of the axions can also be altered. We expect that the momentum of axions at $T=T_*$ is about $m_a$, and
\begin{align}
    y_{k,\delta \theta_{a,{\rm NR}}\gg 1} = \frac{m_a(T_*)}{s_*^{1/3}} = & (Y_a y_k)^{1/4} \frac{m_a^{1/2}(T_*)}{f_a^{1/2}} \nonumber \\
    = & y_k \times \frac{Y_a}{Y_{a,\delta \theta_{a,{\rm NR}}\gg 1}}.
\end{align}
The axions become warmer due to nonlinear evolution.

\subsection{Implications for axion parameter space}
\label{sec:implications}

We now discuss the implication of the new axion dark matter production mechanism, the AMM, for the parameter space of the axion rotation.
We focus on the log-potential model and two field models, where the flatness of the potential of $r$ is naturally understood.
For adiabatic perturbations, the primordial perturbation is given by 
\begin{equation}
    {\cal P}_{\delta_i} =  {\cal P}_{\zeta}.
\end{equation}
We assume a nearly scale-invariant spectrum from the CMB scale to the scale we are interested in and take $ {\cal P}_{\zeta} \simeq 2 \times 10^{-9}$~\cite{Planck:2018vyg} in the following.

\begin{figure}
    \centering   \includegraphics[width=0.495\linewidth]{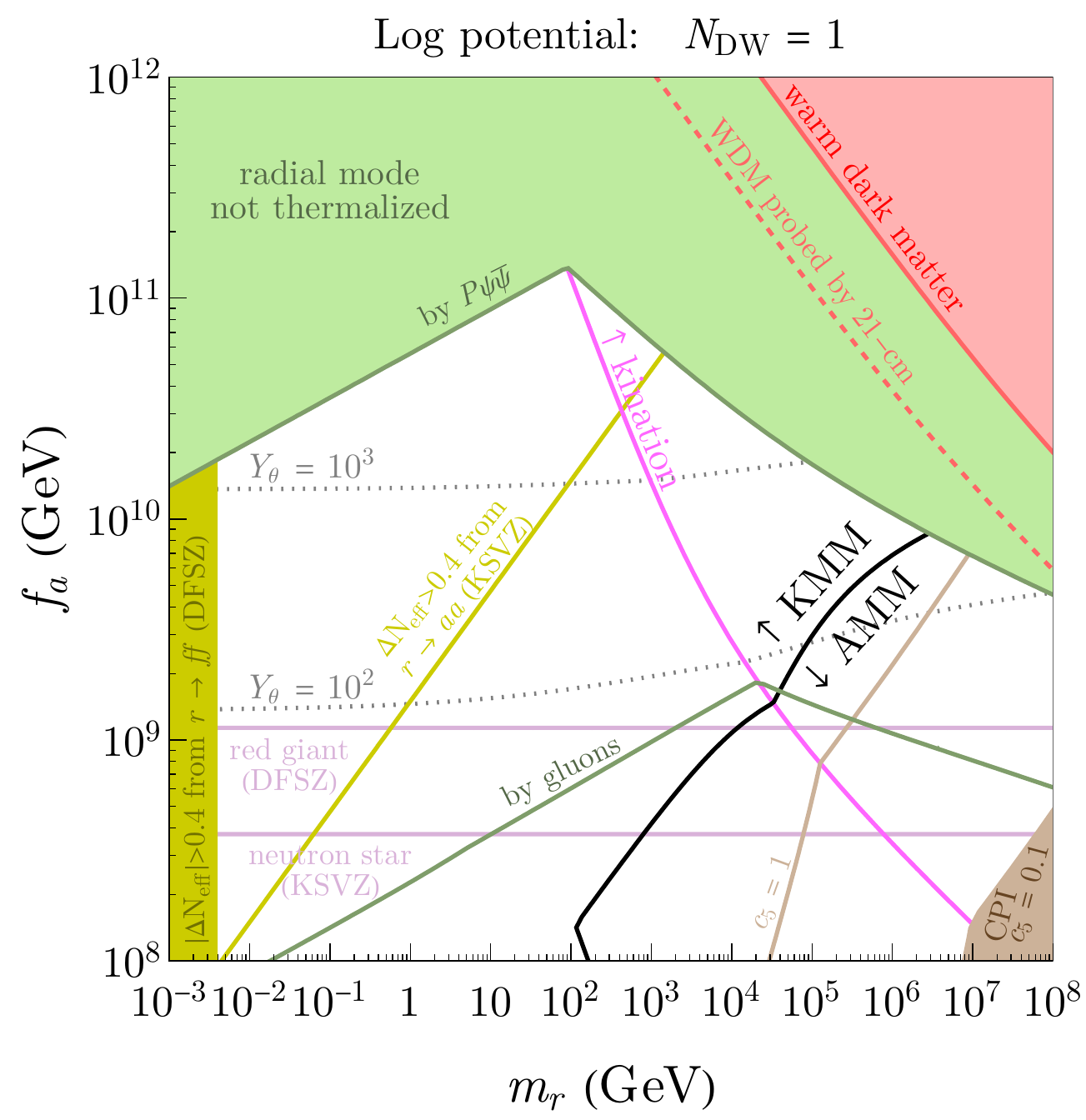}
    \includegraphics[width=0.495\linewidth]{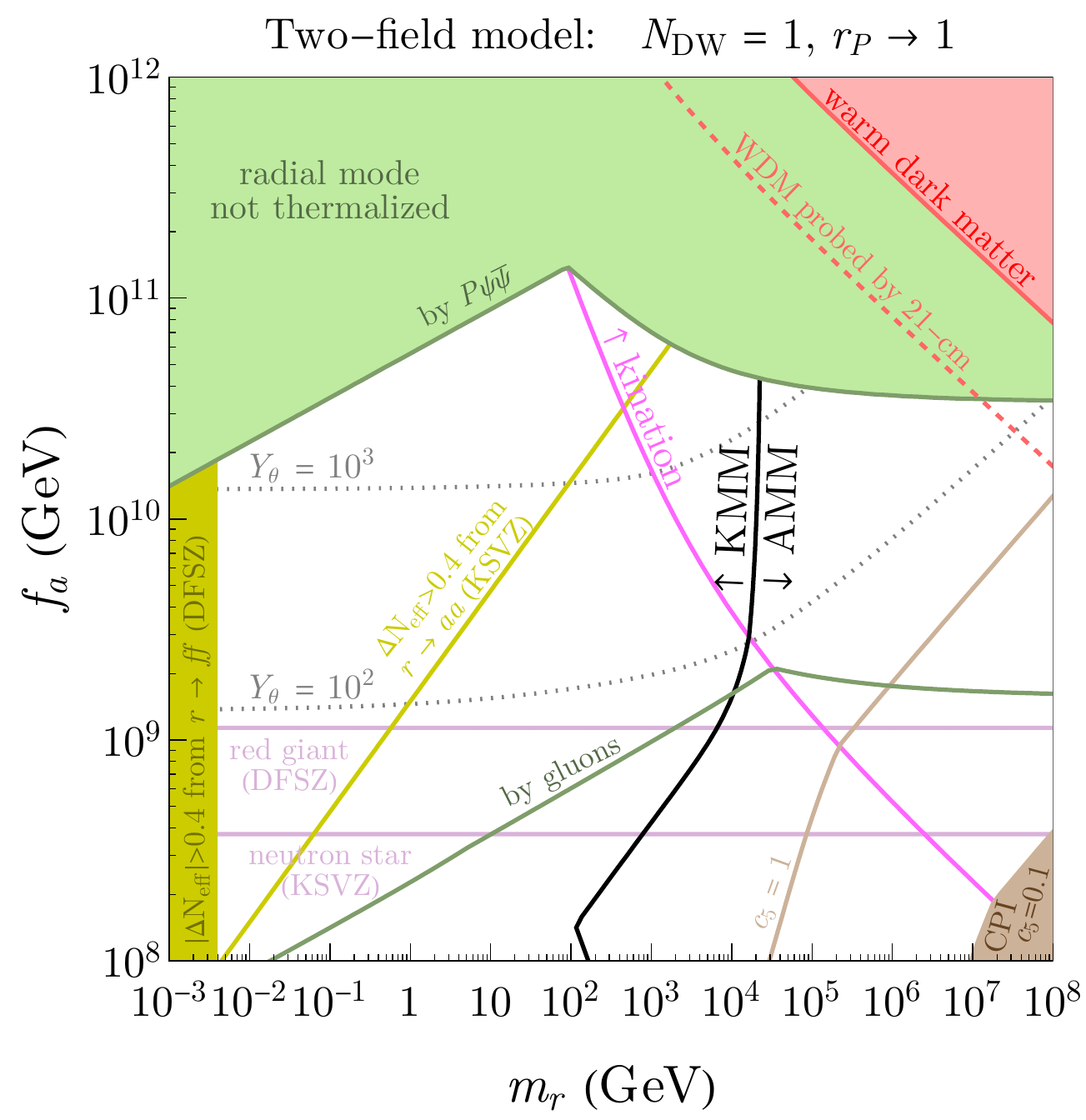}

    \caption{Constraints on the mass of the radial direction $m_r$ and the decay constant $f_a$ for the QCD axion in the log-potential model and the two-field model with $r_P \rightarrow 1$. The gray dotted lines show the contours of the PQ charge yield $Y_\theta$ required to explain the observed dark matter abundance. To the left (right) of the black line, the KMM (the AMM) dominates the production of axion dark matter. To the right of the magenta line, the axion rotation leads to the matter- and kination-dominated eras. The green region (line) shows the upper bound on $f_a$ from the constraint of thermalization of the radial mode via fermion (gluon) scattering processes. The brown region (line) shows the upper bound on $m_r$ from overproduction of baryon asymmetry by chiral plasma instability with $c_5 = 0.1$ ($c_5 = 1$). The red region is excluded because axion dark mater is too warm, while 21-cm observations can probe regions above the red dashed line. The horizontal purple lines show the exclusion by astrophysical constraints on the DFSZ/KSVZ axion. The yellow region (line) is excluded by excessive $\Delta N_{\rm eff}$ from the decay of the thermalized radial mode for the DFSZ (KSVZ) axion. The figure for $r_P \gg 1$ can be found in Appendix~\ref{sec:extra figs}.}
    \label{fig:QCDaxion}
\end{figure}

\begin{figure}
    \centering   \includegraphics[width=0.495\linewidth]{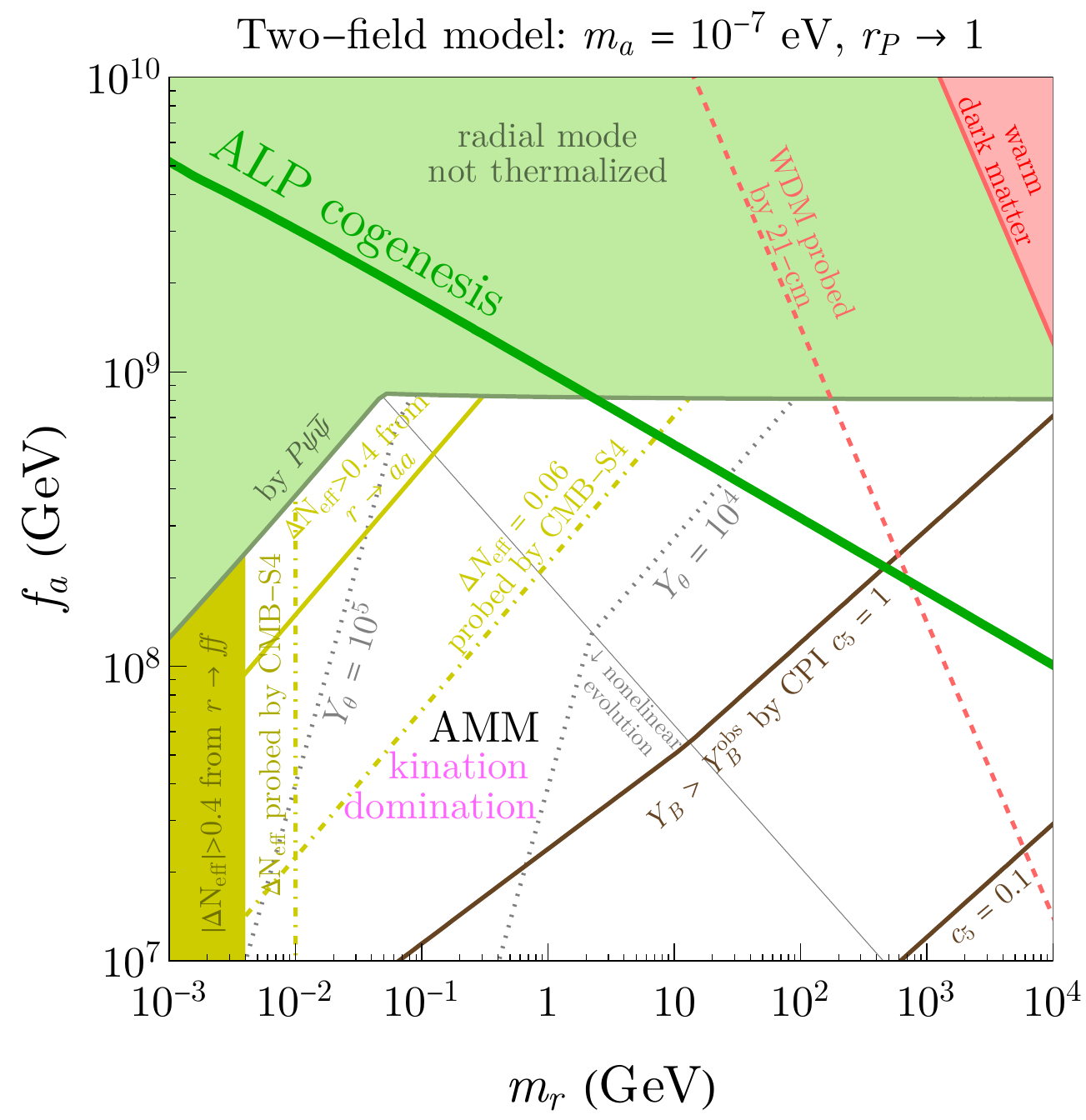}
    \includegraphics[width=0.495\linewidth]{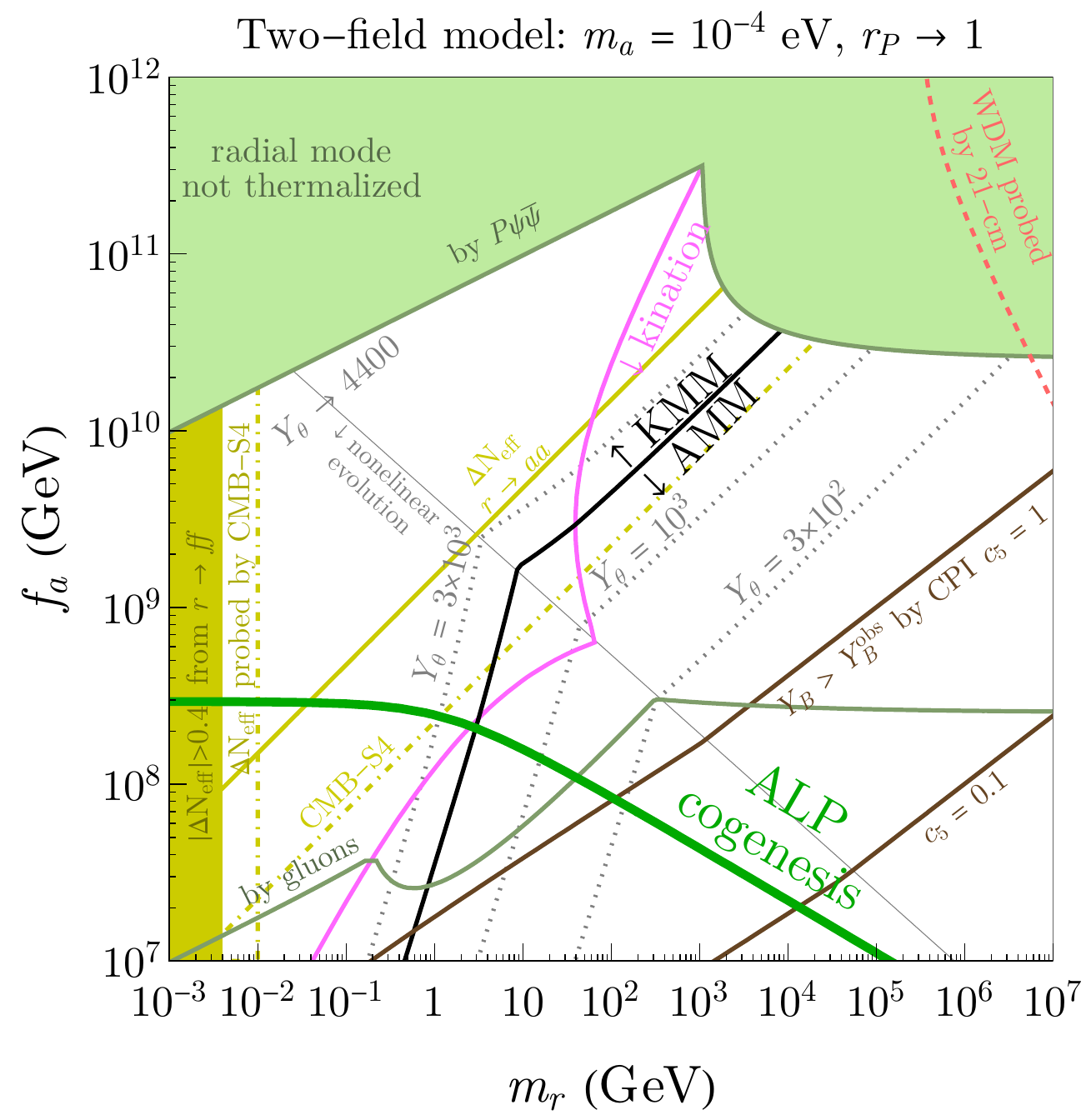}

    \caption{Constraints on the mass of the radial direction $m_r$ and the decay constant $f_a$ for an axion-like-particle with two representative masses in the two-field model with $r_P \rightarrow 1$. Various regions and lines are explained in the caption of Fig.~\ref{fig:QCDaxion}. The constraints for $r_P \gg 1$ and the log-potential model can be found in Appendix~\ref{sec:extra figs}.}
    \label{fig:ALP_rP1}
\end{figure}

Let us first discuss the QCD axion, for which
\begin{equation}
    m_a \simeq 6~{\rm meV} \frac{10^9~{\rm GeV}}{f_a} .
\end{equation}
In Fig.~\ref{fig:QCDaxion}, we show constraints on the parameter space $(m_r,f_a)$ for the log-potential model and the two-field model with $r_P\rightarrow1$. Here we require that the sum of the KMM and AMM contributions explain the observed dark matter abundance.
To the right of the black solid line, the AMM contribution dominates over the KMM.
The kink at the bottom part of the line is due to $\delta \theta_{a,{\rm NR}}>1$ that modifies the AMM contribution.
The dotted contours show the required $Y_\theta$. 
To the right of the magenta solid line, the axion rotation dominates the universe and hence there exists a kination-dominated era. 
Inside the green-shaded region, the thermalization is not efficient enough to obtain the required $Y_\theta$ even if the scattering is via Yukawa couplings.
The positively/negatively sloped boundary is determined by Eqs.~\eqref{eq:Y_max_th} and \eqref{eq:Y_max_th_dom} with $b =0.1$, respectively. 
The green solid line shows the constraint when the thermalization occurs via a coupling with gluons. Inside the red-shaded region, the axion dark matter is too warm. Future measurements of the 21 cm lines can probe above the red dotted line~\cite{Sitwell:2013fpa}.
Above the yellow line, a thermal abundance of the radial direction $r$ produced by the thermalization of the rotation decays into axions and creates too much dark radiation $\Delta N_{\rm eff} > 0.4$~\cite{Planck:2018vyg}. If $r$ couples to the electrons, as is the case in the DFSZ model, $r$ can be in thermal equilibrium with the SM bath when it becomes non-relativistic and its energy density can be exponentially suppressed. Still, it can heat up the electrons and photons after neutrinos decouple to create negative $\Delta N_{\rm eff}$, excluding the yellow-shaded region~\cite{Ibe:2021fed}. Below the horizontal light purple lines are constrained by the red-giant~\cite{Capozzi:2020cbu,Straniero:2020iyi} and neutron-star~\cite{Leinson:2021ety,Buschmann:2021juv} cooling bounds for the DFSZ and KSVZ models, respectively.
The bounds can be relaxed in models where the couplings of the axion with nucleons and electrons are suppressed~\cite{DiLuzio:2017ogq,Bjorkeroth:2019jtx, Badziak:2023fsc,Takahashi:2023vhv}.
The brown-shaded region will be explained in Sec.~\ref{sec:axiogenesis}.

We next discuss an axion-like particle (ALP), where $m_a$ and $f_a$ are independent free parameters. In Fig.~\ref{fig:ALP_rP1}, we show the constraint on $m_r$ and $f_a$ for the two-field model with $r_P\rightarrow 1$ and two representative masses of the ALP. The meanings of the shadings and lines are the same as those of Fig.~\ref{fig:QCDaxion}. In the left panel, the thermalization by a coupling with gauge bosons is not effective enough inside the plotted region. The thick green line with a label ``ALP cogenesis" will be explained in Sec.~\ref{sec:axiogenesis}.

\begin{figure}
    \centering
    \includegraphics[width=0.49\linewidth]{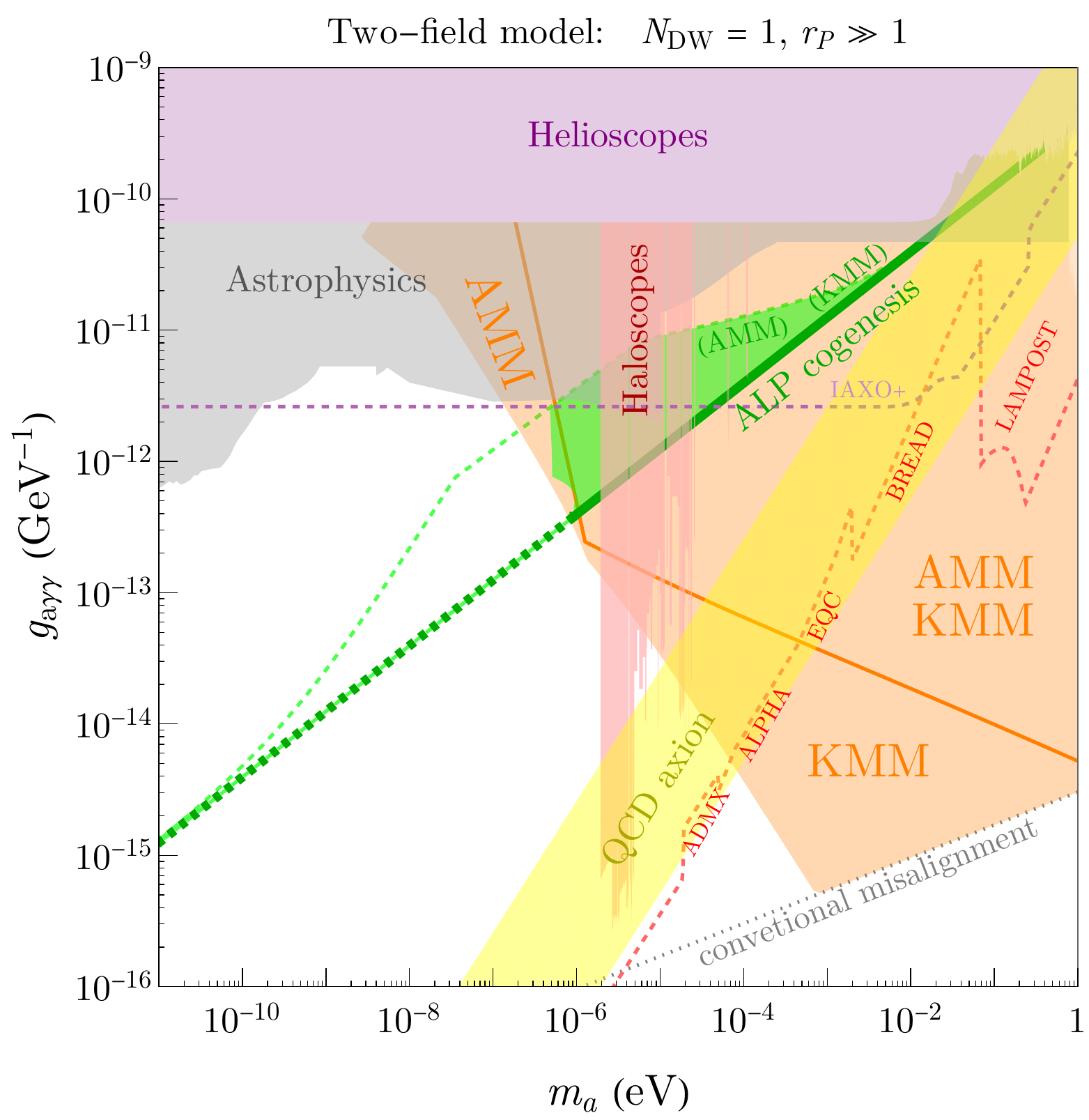}
    \includegraphics[width=0.49\linewidth]{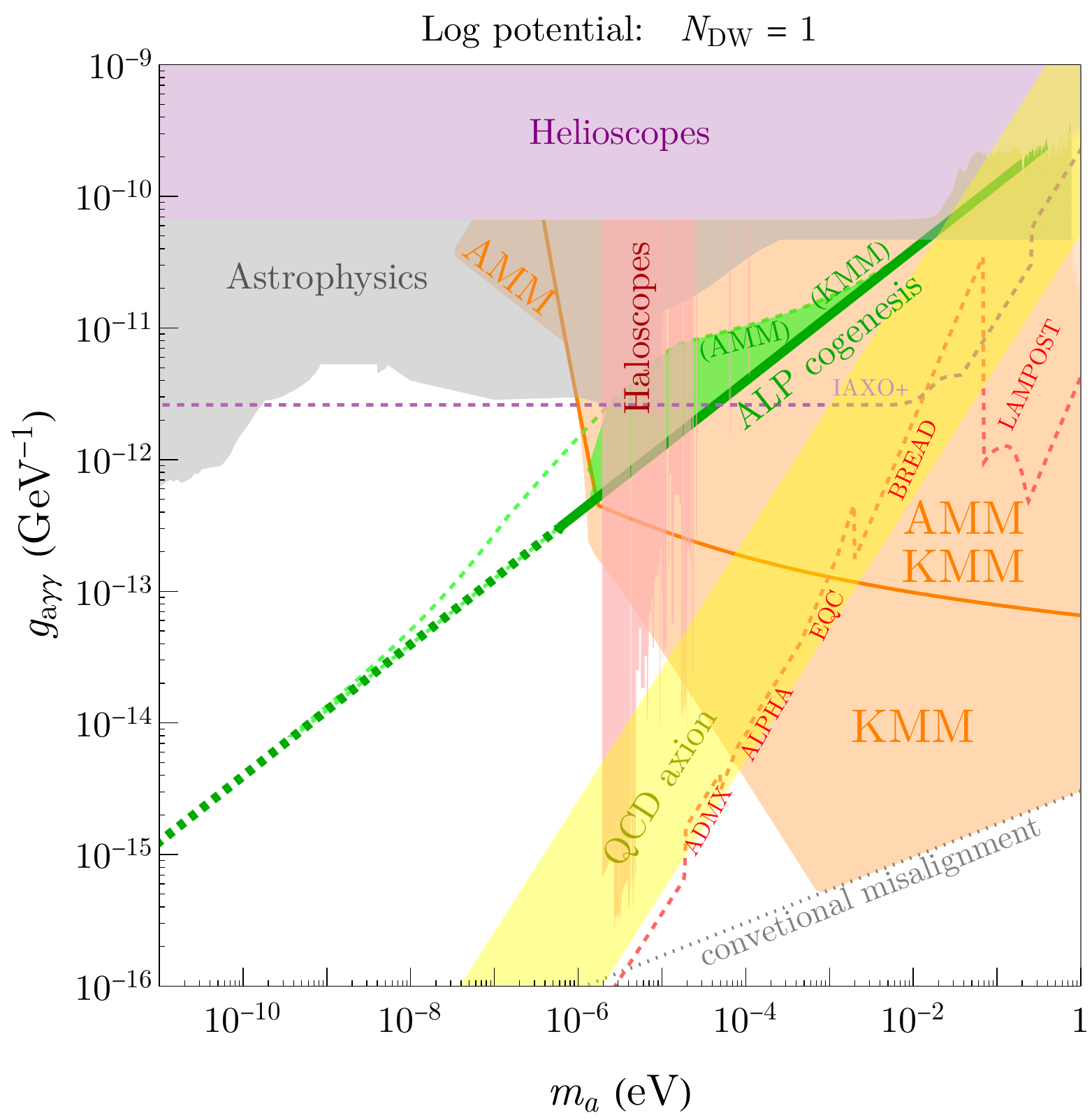}
    \caption{Same as Fig.~\ref{fig:AMM_cogen_rP1} but for the two-field model with $r_P\gg 1$ and the log-potential model.}
    \label{fig:AMM_cogen2}
\end{figure}

The orange-shaded regions in Figs.~\ref{fig:AMM_cogen_rP1} and~\ref{fig:AMM_cogen2} show the parameter regions in $(m_a,g_{a\gamma\gamma})$ where the AMM or KMM can explain the observed dark matter abundance and there exists some range of $m_r$ that satisfies all constraints.  Here we assume $g_{a\gamma\gamma} =\alpha/(2\pi f_a)$. The left boundary of the orange-shaded region is determined by the thermalization constraint.
In the regions labeled AMM or KMM, the AMM or KMM contribution dominates over the other one, respectively. In the regions with both labels, either can dominate depending on $m_r$. Below the gray dotted line, the conventional misalignment mechanism overproduces axion dark matter assuming an initial misalignment angle of order unity.
The green line and the green-shaded region will be explained in Sec.~\ref{sec:axiogenesis}.

In the gravity-mediated supersymmetry breaking scenario, the mass of the PQ symmetry-breaking field $m_r$ is expected to be of the same order as those of the scalar particles in supersymmetric Standard Models. Then $m_r$ is expected to be above the TeV scale, for which the AMM tends to dominate over the KMM both for the QCD axion and ALPs. Furthermore, in the minimal supersymmetric Standard Model, the observed Higgs mass of 125 GeV requires that the scalar mass scale is larger than $O(10)$/$O(100)$ TeV for ${\rm tan}\beta\gg 1$/${\rm tan}\beta = O(1)$~\cite{Okada:1990gg,Okada:1990vk,Ellis:1990nz,Haber:1990aw}, for which the AMM is more likely to dominate.
In the gauge-mediated supersymmetry breaking scenario, on the other hand, $m_r$ may be much below the TeV scale and the KMM may dominate.

\subsection{Implications for axiogenesis scenarios}
\label{sec:axiogenesis}

The AMM has implications for baryogenesis by axion rotation (axiogenesis).
The angular momentum of the axion rotation, namely, the PQ charge, can be converted into the asymmetry of particles in the thermal bath, and their chemical potentials are $O(0.1) \dot{\theta}_a$. With baryon number violation, those asymmetries can be converted into baryon asymmetry~\cite{Co:2019wyp}. 

\subsubsection{Minimal axiogenesis}

Due to the efficient baryon number violation by the weak sphaleron process, the baryon number density is given by the thermal equilibrium value $c_B\dot{\theta_a}T^2$, where $c_B$ a model-dependent coefficient that is typically $O(0.1)$~\cite{Co:2019wyp}.
The baryon asymmetry is frozen at the electroweak phase transition and is given by
\begin{align}
    \frac{n_B}{s} = c_B Y_\theta \times \frac{T_{\rm sp}^2}{f_a^2 } \simeq 8\times 10^{-11} \frac{c_B}{0.05} \left(\frac{10^9~{\rm GeV}}{f_a} \right)^2 \left(\frac{T_{\rm sp}}{130~{\rm GeV}}\right)^2 \frac{Y_\theta}{10^5},
\end{align}
where $T_{\rm sp}$ is the temperature below which the electroweak sphaleron process becomes inefficient, which is about 130 GeV in the SM~\cite{DOnofrio:2014rug}.
For the QCD axion, if the parameters are fixed such that the observed baryon abundance is explained by axion rotation, the KMM overproduces axion dark matter unless $f_a \lesssim 10^{7}$ GeV, the electroweak phase transition temperature is raised, or the coupling of the QCD axion to gluons is much weaker than that to another SM particle. 
Extra contributions from the AMM only make this minimal axiogenesis scenario even harder to realize.

For ALPs, on the other hand, we may explain the observed baryon asymmetry without overproducing axion dark matter. On the thick green line in Fig.~\ref{fig:ALP_rP1}, such a cogenesis scenario is possible. Here we take $c_B=0.1$.
Assuming that the KMM dominates over the AMM, the prediction of cogenesis is~\cite{Co:2020xlh}
\begin{align}
     f_a \simeq 2 \times 10^9~{\rm GeV} 
    \left(\frac{c_B}{0.1}\right)^{{ \scalebox{1.01}{$\frac{1}{2}$} }} 
    \left( \frac{\mu{\rm eV}}{m_a} \right)^{ \scalebox{1.01}{$\frac{1}{2}$} } 
    \left( \frac{T_{\rm sp}}{130~{\rm GeV}} \right),
\end{align}
which corresponds to the horizontal segment of the green line. Once the AMM dominates, smaller $f_a$ is predicted.

The prediction on the axion-photon coupling when the KMM dominates is shown by the green line in Figs.~\ref{fig:AMM_cogen_rP1} and \ref{fig:AMM_cogen2}. Here the coupling is given by
\begin{align}
{\cal L} &= -\left(\frac{g_{a\gamma\gamma}}{4} \right) a \,  \epsilon^{\mu\nu\rho\sigma}F_{\mu\nu}F_{\rho\sigma} ,
\end{align}
and we assume $g_{a\gamma\gamma} =\alpha/(2\pi f_a)$. In the dotted segment, thermalization of the rotation by the process discussed in Sec.~\ref{sec:rotation} with $b=0.1$ makes it impossible to obtain the required $Y_\theta$ and a more efficient thermalization mechanism is required.
In the green-shaded region, the AMM dominates the production of axion dark matter while explaining the observed baryon asymmetry.
Fig.~\ref{fig:cogenesis} provides a zoomed-in view of the cogenesis region.
The blue dot-dashed lines show the required $m_r$. For a fixed $m_r$, as $m_a$ increases, the AMM becomes subdominant and the prediction converges to that of cogenesis by the KMM.
Above the green-shaded region, the process discussed in Sec.~\ref{sec:CPI} overproduces baryon asymmetry. To the left of the green-shaded region, thermalization of the rotation to obtain the required $Y_\theta$ by the process discussed in Sec.~\ref{sec:rotation} is impossible. 
The predicted parameter region can be probed by proposed experiments.
If there exists a more efficient thermalization mechanism, the green-shaded region can be extended to the region bounded by the green dashed line.

\subsubsection{Magneto-axiogenesis}
\label{sec:CPI}

\begin{figure}
    \centering   
    \includegraphics[width=0.495\linewidth]{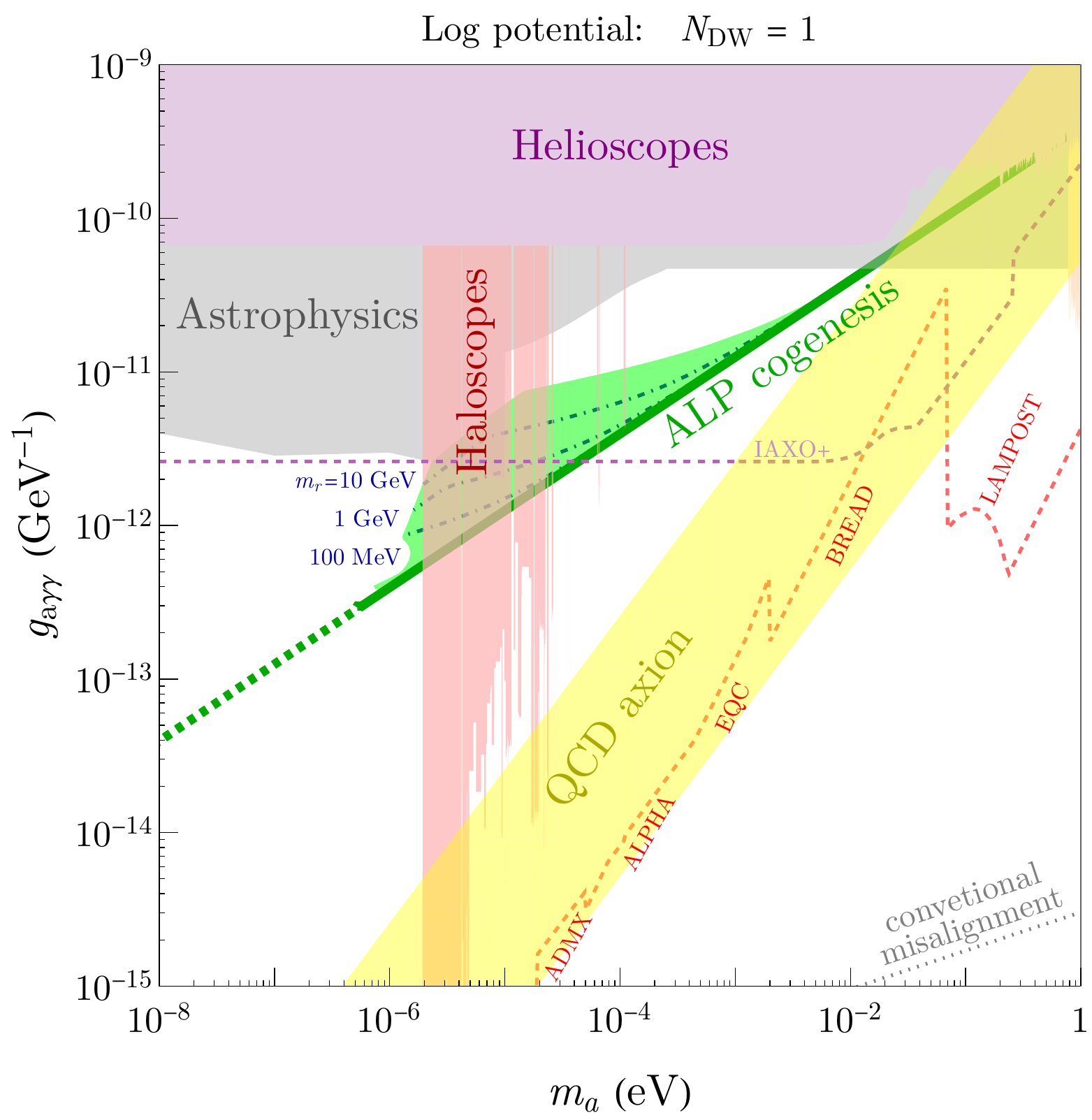}
    \includegraphics[width=0.495\linewidth]{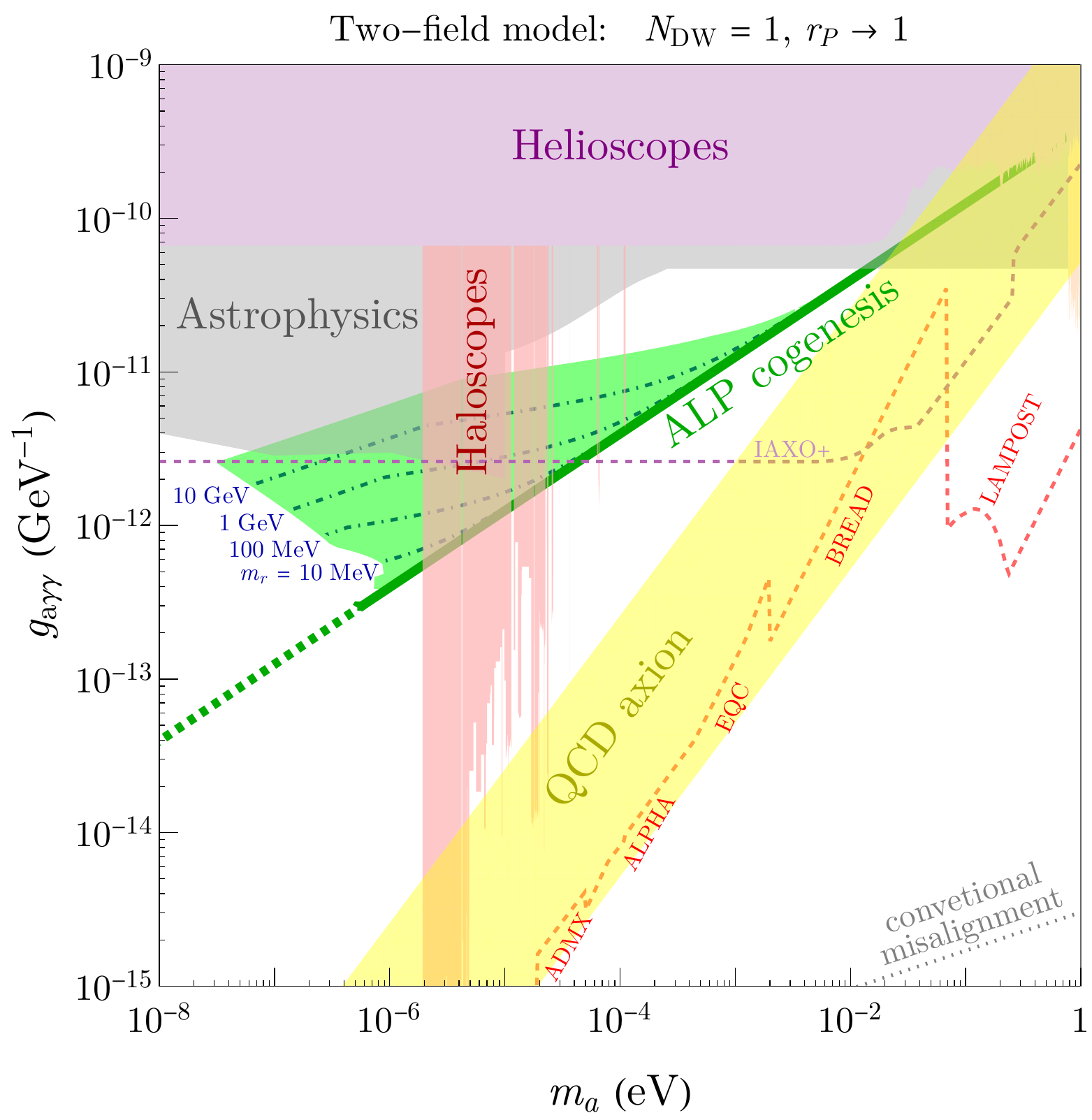}
    \caption{Implications of ALP cogenesis for axion searches in the log-potential model and the two-field model with $r_P \rightarrow 1$ and $\ndw = 1$. The figure for $r_P \gg 1$ can be found in Appendix~\ref{sec:extra figs}. In the green-shaded region, dark matter and baryon asymmetry of the universe can be explained by the rotation of an ALP field in the field space.}
    \label{fig:cogenesis}
\end{figure}

The rotation can also produce baryon asymmetry by the generation of hypermagnetic fields and the electroweak anomaly of the baryon symmetry. The rotation, through its direct coupling with the hypercharge gauge field or through the generation of asymmetry of hypercharged fermions in the thermal bath, induces chiral plasma instability (CPI) of the hypercharge gauge field~\cite{Turner:1987bw,Joyce:1997uy}, producing helical hypermagnetic fields at a rate~\cite{Kamada:2018tcs,Domcke:2019mnd,Co:2022kul}
\begin{equation}
    \Gamma_{\rm CPI} \simeq \frac{\alpha_Y^2 c_5^2 \dot{\theta}^2}{100 \pi^2 T},
\end{equation}
where $\alpha_Y$ is the hypercharge fine structure constant and $c_5$ is a model- and temperature-dependent constant that is typically $O(0.01-1)$ for the QCD axion and $O(1)$ for ALPs that couple to photons~\cite{Co:2022kul}.%
\footnote{The possibility of small $c_5$ for the QCD axion comes from cancellation in $SU(5)$ limit. If an ALP does not have anomalous couplings to gauge bosons and has couplings with fermions smaller than $1/f_a$-suppressed ones, $c_5$ can be much smaller than 1.}
Via the inverse cascade process, the correlation length of the helical hypermagnetic fields becomes much longer than the typical length scale of the thermal bath $\sim T^{-1}$, and the helicity density is conserved without being washed out by the interaction with the thermal bath~\cite{Brandenburg:2017rcb}. 
The helical hypermagnetic field is converted into a helical magnetic field around the electroweak phase transition, and the weak and hypercharge anomaly of baryon symmetry produces baryon asymmetry~\cite{Giovannini:1997gp,Giovannini:1997eg,Kamada:2016eeb,Kamada:2016cnb}.
Note that $\Gamma_{\rm CPI}/H$ is maximized at MK.
If $\Gamma_{\rm CPI} > O(10) H$ at MK, the CPI becomes fully effective, and the resultant baryon asymmetry is~\cite{Co:2022kul}
\begin{equation}
    \frac{n_B}{s} \sim c_B^{\rm dec} Y_\theta \frac{T_{\rm MK}^2}{f_a^2},  
\end{equation}
where $c_B^{\rm dec}$ is the conversion efficiency factor whose value is uncertain, but is $O(10^{-3}-1)$~\cite{Jimenez:2017cdr}.
We find that the resultant baryon asymmetry is always larger than the observed baryon asymmetry when the AMM or KMM explains the observed dark matter abundance. We thus require that the CPI does not become fully effective, i.e., $\Gamma_{\rm CPT} < 10 H$ at MK. The constraint is shown by brown-shaded regions and brown lines in Figs.~\ref{fig:QCDaxion} and~\ref{fig:ALP_rP1}, which exclude large $m_r$. Here we take $c_5=0.1$ and $1$ as reference values for the brown-shaded regions and brown lines, respectively.  Around the boundary of the constrained region, the CPI can be marginally efficient and the observed baryon asymmetry can be explained at the expense of tuning the parameters of the theory.
In such a parameter region, the AMM contribution dominates over the KMM contribution.
The upper boundary of the green-shaded regions in
Figs.~\ref{fig:AMM_cogen_rP1}, \ref{fig:AMM_cogen2}, and \ref{fig:cogenesis} are determined by the CPI bound with $c_5=1$.

\subsubsection{Lepto-axiogenesis}

In lepto-axiogenesis, $B-L$ is produced by the dimension-five Majorana neutrino mass operator~\cite{Co:2020jtv,Kawamura:2021xpu,Barnes:2022ren}. The baryon number produced per Hubble time is given by
\begin{equation}
    \frac{\Delta n_B}{s}\simeq \frac{45}{2\pi^2 g_*(T)} C(T) \frac{m_\nu^2}{v_{\rm EW}^4} \frac{\dot{\theta} T^2}{H^2},
\end{equation}
where $C (T)$ is a model-dependent constant that is $O(10^{-2}-10^{-3})$. See~\cite{Barnes:2022ren} for the computation and exact values of $C$.  
From the scaling of $\dot{\theta}$, $H$, and $T$, one can see that the production is dominated before RM, for which $\dot{\theta} = m_r$ and the baryon asymmetry produced per Hubble time is given by
\begin{equation}
    \frac{\Delta n_B}{s}\simeq 8\times 10^{-11} \frac{m_r}{100~{\rm TeV}} \left(\frac{m_\nu}{50~{\rm meV}}\right)^2 \frac{C}{0.02} \left(\frac{106.75}{g_*}\right)^{3/2}.
\end{equation}
From this equation, one can see that we need $m_r > O(10)$ TeV to explain the observed baryon asymmetry. Figs.~\ref{fig:QCDaxion} and \ref{fig:ALP_rP1} show that the AMM dominates over the KMM for such a large mass. 
When the right-handed neutrinos are light, they may be in the thermal bath while the axion rotates in the early universe, and the production becomes more efficient. $m_r$ may be as low as $O(1)$ MeV~\cite{Barnes:2024jap}, for which the KMM can dominate.

\vspace{0.5cm}

There are other axiogenesis scenarios, including those with R-parity violation in supersymmetric theories~\cite{Co:2021qgl} and a sphaleron process of a new gauge interaction~\cite{Harigaya:2021txz}. The viable parameter space of those scenarios will be also affected by the AMM. We leave the investigation of those scenarios for future work.

\section{Summary and discussion}
\label{sec:summary}

Axion rotation is a Bose-Einstein condensation of PQ charges with repulsive self-interaction, i.e., superfluid, and the fluctuations around the rotating background have a sound-wave mode, which may be identified with axion fluctuations once the PQ charges are diluted by cosmic expansion.
In this paper, we follow the cosmological evolution of the fluctuations in axion rotation and formulate the computation of the dark matter abundance originating from the sound-wave mode. We refer to this method of axion dark matter production as the ``acoustic misalignment mechanism'' (AMM). 
We identify the parameter space where the AMM contribution dominates over that from the kinetic misalignment mechanism, which occurs for a sufficiently large mass of the radial direction of the PQ breaking field. The axion dark matter may be warm enough to affect structure formation.

The results have rich implications for axion searches and the cosmology of axion rotation.
The parameter region resulting in kination dominance shrinks. The cogenesis of axion dark matter and baryon asymmetry through the axion rotation and the electroweak sphaleron process, i.e., minimal axiogenesis, predicts larger axion couplings to the SM particles in comparison with the case where the AMM contribution is subdominant.
Baryogenesis by axion rotation with the aid of Majorana neutrino mass terms, the so-called lepto-axiogenesis scenario, requires that the mass of the radial direction is above $O(10)$ TeV, and for such a large mass, 
the AMM contribution dominates over the kinetic misalignment mechanism.
We leave the investigation of implications for other baryogenesis scenarios for future work.

We have discussed the implications for the parameter space assuming that the perturbations are adiabatic and nearly scale invariant. Our computation is applicable to the case where the adiabatic perturbation has large or small amplitudes at small scales or the axion rotation has isocurvature fluctuations. Indeed, if the angular direction is nearly massless during inflation, quantum fluctuations of it are generated and the PQ charge density is modulated. The isocurvature perturbations, however, lead to the growth of field perturbations even outside the horizon and can create large boundary-less domain walls~\cite{Co:2020dya}, which are stable even for $\ndw =1$. The enhanced field perturbations also produce too large dark matter isocurvature perturbations from the misalignment mechanism~\cite{Co:2022qpr}.
To avoid the domain wall and isocurvature problems, the isocurvature perturbations must be highly blue-tilted, which can be achieved by the dynamics of the radial direction during inflation~\cite{Kasuya:2009up}.

In this work, we have focused on the scenario where the PQ symmetry-breaking field itself undergoes the rotation initiated by the Affleck-Dine mechanism, but our results apply to the case where another scalar field initially rotates and its charges are transferred into the PQ symmetry-breaking field~\cite{Domcke:2022wpb}. The other field obtains charge density fluctuations, and the fluctuations are transferred into those of the PQ symmetry-breaking field, which become axions after enough redshift of the charges.

In the AMM, axion dark matter have large fluctuations at small scales, which  may result in small gravitationally bound dark-matter halos called axion minihalos (or miniclusters)~\cite{Hogan:1988mp,Kolb:1993zz,Kolb:1993hw,Kolb:1994fi} or self-interaction bound solitonic objects called axion stars~\cite{Tkachev:1991ka}. We leave the investigation of this possibility for future work. 

The formulation developed in this paper can be applied to a broader class of models beyond dark matter production. If the axion mass is negligible, the model is constrained by the overproduction of dark radiation. The possible enhancement of the perturbations during the pre-kination phase gives a stronger constraint than the model-independent bound derived in~\cite{Eroncel:2025bcb}, which analyzes the evolution of perturbations during the kination phase alone. 

Our results demonstrate that cosmic perturbations of generic fields following the kination equation of state can produce dark matter, and the abundance arising from the kination phase can be computed using our formulation, although the possible nonlinear evolution when fluctuations become non-relativistic may be model dependent. Additionally, contributions from the pre-kination phase may require more careful investigation.

Our results highlight the significant role of the AMM in axion dark matter production and its cosmological implications. The interplay between axion rotation, its perturbations, subsequent evolution, and baryon-number violation provides a solid framework for addressing both dark matter and baryon asymmetry. Furthermore, our results suggest new experimental targets for axion searches, as the AMM generally predicts larger axion couplings to Standard Model particles. Future studies should explore the nonlinear evolution of large perturbations, the impact of isocurvature fluctuations, and potential observational signatures in structure formation. These directions will refine our understanding of axion cosmology and the observational testability of the AMM scenarios.

\vspace{0.5cm}

{\bf Note added:} As this paper was being completed we became aware of overlapping work in preparation from another group~\cite{Eroncel:2025}.

\section*{Acknowledgment}
RC, AG, and KH thank Nicolas Fernandez and Jessie Shelton for collaboration on a related project. 
This work was supported by DOE grant DE-SC-0013642 (AB and LTW) and DE-SC0025242 (KH) at the University of Chicago, DE-SC0025611 (RC) at Indiana University, and the GRASP initiative at Harvard University (AG); a Grant-in-Aid for Scientific Research from the Ministry of Education, Culture, Sports, Science, and Technology (MEXT), Japan 20H01895 (KH); and by World Premier International Research Center Initiative (WPI), MEXT, Japan, Kavli IPMU (KH). In addition, AB is also supported by the Fermi Forward Discovery Group, LLC under Contract No. 89243024CSC000002 with the U.S. Department of Energy, Office of Science, Office of High Energy Physics.

\appendix

\section{Phonon dispersion relation}
\label{app:phonon}

In this appendix, we compute the dispersion relation of the perturbations around rotating field $P$. We consider the case where $P$ has a canonical kinetic term. The equation motions of $r$ and $\theta$ is
\begin{align}
    \ddot{r} - r \dot{\theta}^2 - \partial_i^2 r + r (\partial_i \theta)^2 + V_r =0,~~
    r \ddot{\theta} + 2 \dot{r}\dot{\theta}- 2 \partial_i r \partial_i \theta - r \partial_i^2 \theta = 0.
\end{align}
The equation of motion of the zero mode $r_0(t)$, $\theta_0(t)$  are
\begin{align}
    \ddot{r_0} - r_0 \dot{\theta}^2_0 + V_r(r_0)=0, ~~
    r_0 \ddot{\theta}_0 + 2 \dot{r}_0\dot{\theta}_0 =0.
\end{align}
For a circular motion, where $\dot{r}_0=0$ and $\ddot{r}=0$, the solution is given by
\begin{align}
    \dot{\theta}_0^2 = \frac{1}{r_0}V_r(r_0),~~r_0^2 \dot{\theta_0} = n_\theta =~\text{constant}.
\end{align}
The first-order perturbations $\delta r$, $\delta \theta$ follows
\begin{align}
    \ddot{\delta r} - 2 r_0 \dot{\theta}_0 \dot{\delta \theta} + (V_{rr} - \partial_i^2 - \dot{\theta}_0^2)\delta r = 0 ,\\
    \ddot{\delta \theta} +  2 \dot{\theta_0} \frac{\dot{\delta r}}{r_0} - \partial_i^2\delta\theta =0.
\end{align}
In the Fourier space, they are
\begin{equation}
    \begin{pmatrix}
    E^2 - k^2 + \dot{\theta}_0^2 - V_{rr} &  - 2 i E \dot{\theta}_0 \\
     2 i E \dot{\theta}_0 & E^2 - k^2
    \end{pmatrix}
    \begin{pmatrix}
        \delta r_k \\ r_0 \delta\theta_k
    \end{pmatrix} =0 .
\end{equation}
Non-trivial solutions exists when the determinant of the matrix is zero,
\begin{equation}
    E^4 - E^2 \left(2k^2 + 3 \dot{\theta}_0^2 + V_{rr}   \right) + k^2 \left( k^2 - \dot{\theta}^2_0 + V_{rr} \right) =0.
\end{equation}
The two branches of solutions for $E$ is
\begin{equation}
    E^2 = \frac{1}{2}\left(
    2 k^2 + 3 \dot{\theta}_0^2 + V_{rr} \pm \sqrt{ 16k^2 \dot{\theta}_0^2 + \left( V_{rr} +3\dot{\theta}_0^2  \right)^2 } 
    \right).
\end{equation}
For low $k$, at the leading order, they are
\begin{align}
    E ^2 & \simeq \frac{V_{rr} - \dot{\theta}_0^2}{ V_{rr}+ 3 \dot{\theta}_0^2}k^2 = \frac{V_{rr} - V_r/r}{ V_{rr}+ 3 V_r/r}k^2 \equiv c_s^2 k^2, \\
    E^2 & \simeq 3 \dot{\theta}^2 + V_{rr} = 3  V_r/r + V_{rr}.
\end{align}
The first one is gapless and corresponds to phonon modes. $c_s$ coincides with Eq.~\eqref{eq:cs} derived for cosmological perturbations. The second one has an energy gap and corresponds to the excitation of the radial mode.

\section{Axion parameter space}
\label{sec:extra figs}

\begin{figure}[t]
    \centering   
\includegraphics[width=0.495\linewidth]{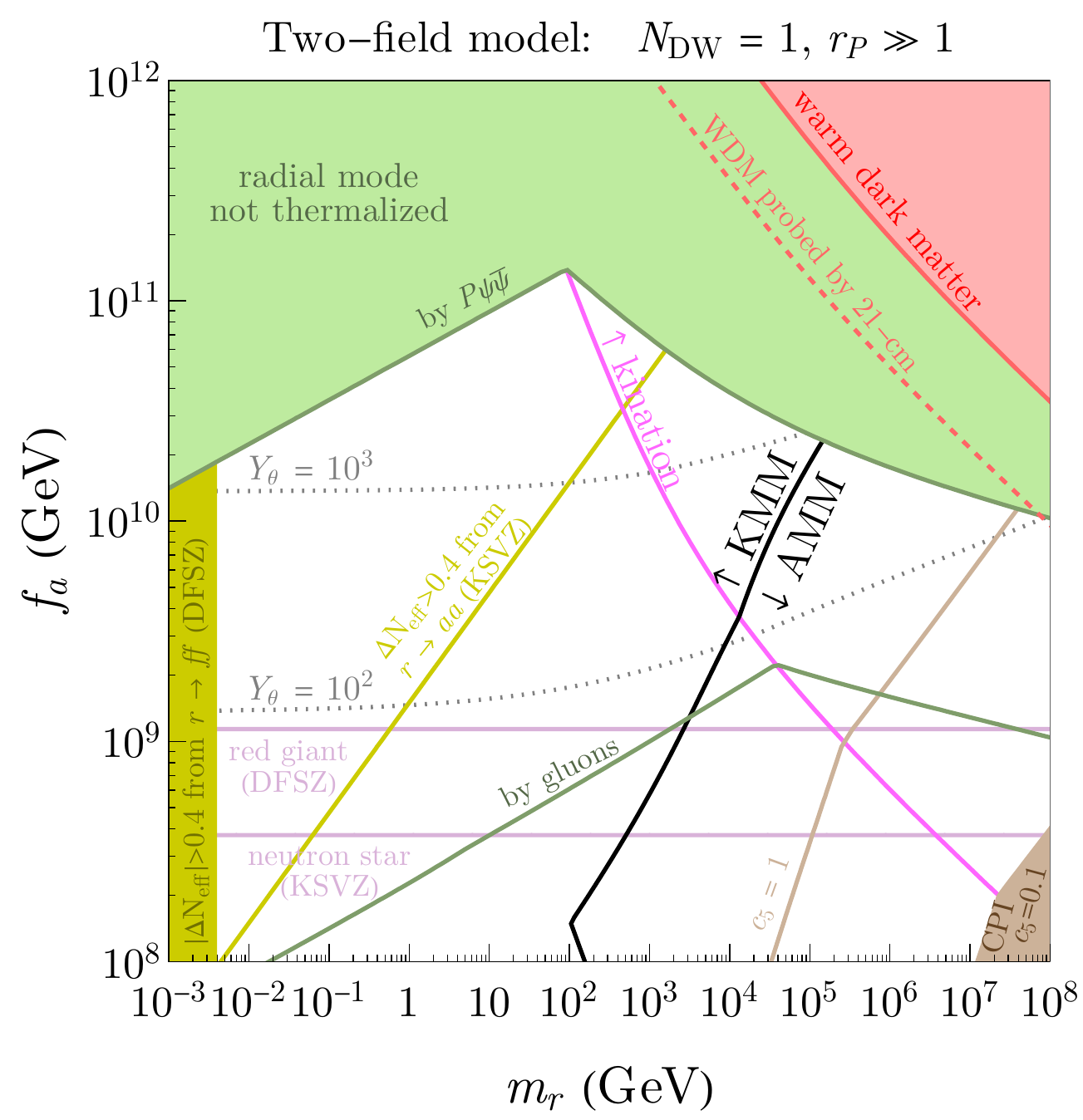}
    \caption{Same as Fig.~\ref{fig:QCDaxion} but for $r_P \gg 1$.}
    \label{fig:QCDaxion2}
\end{figure}

\begin{figure}[t]
    \centering   \includegraphics[width=0.495\linewidth]{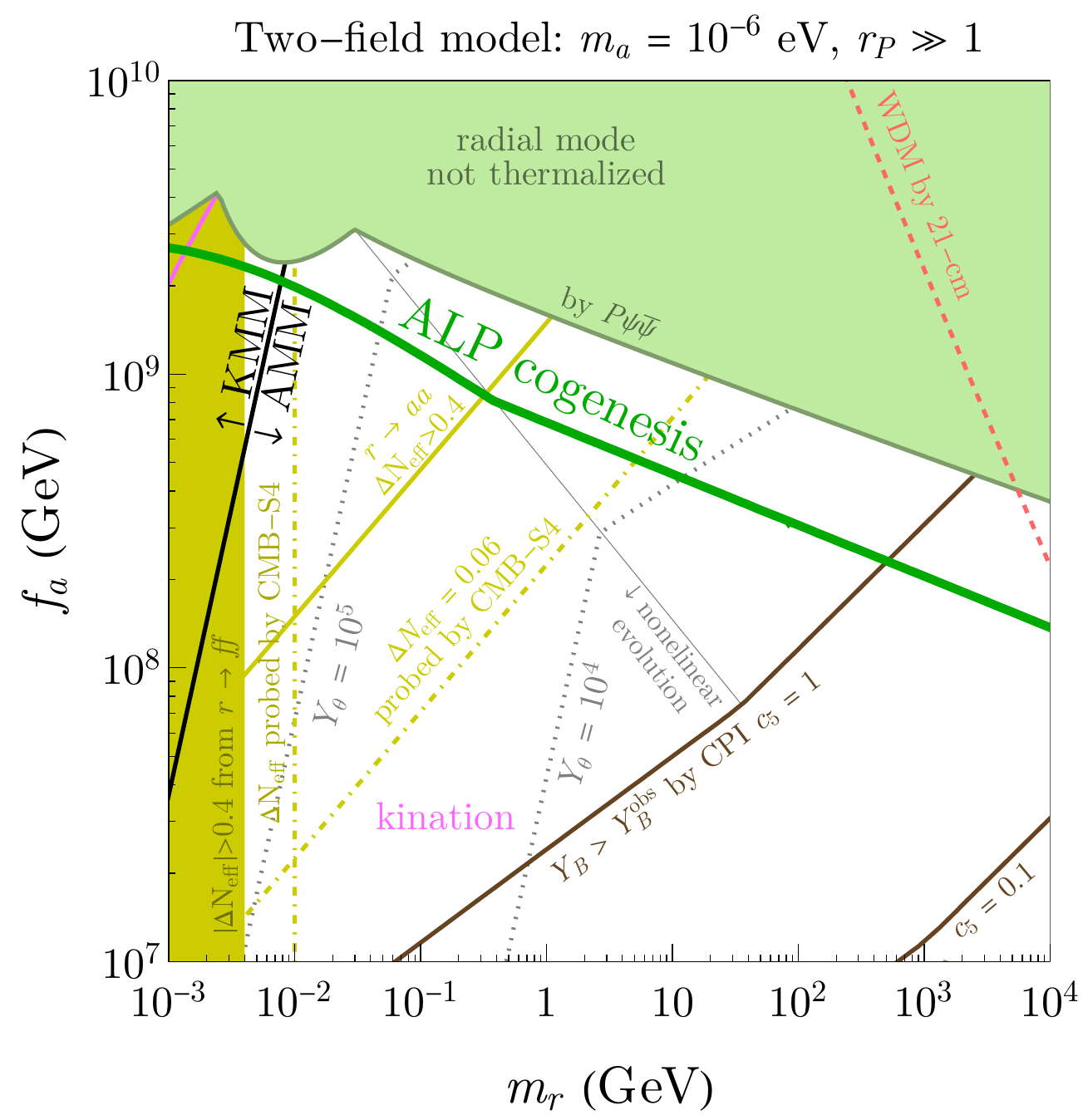}
    \includegraphics[width=0.495\linewidth]{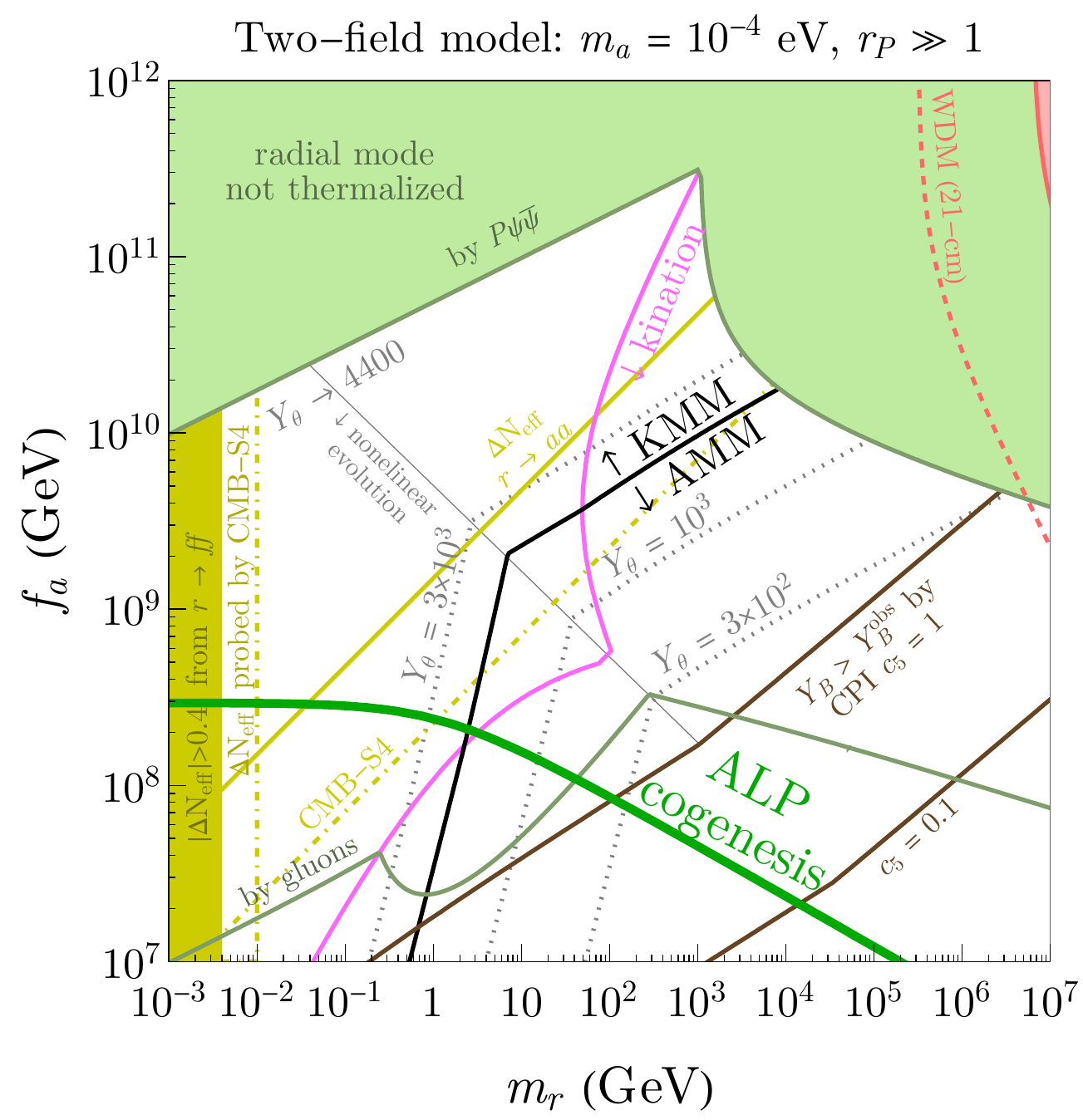}

    \caption{Same as Fig.~\ref{fig:ALP_rP1} but for $r_P \gg 1$, with $m_a = 10^{-6} \eV$ ($10^{-4} \eV$) in the left (right) panel.}
    \label{fig:ALP_rPlarge}
\end{figure}

\begin{figure}[t]
    \centering   \includegraphics[width=0.495\linewidth]{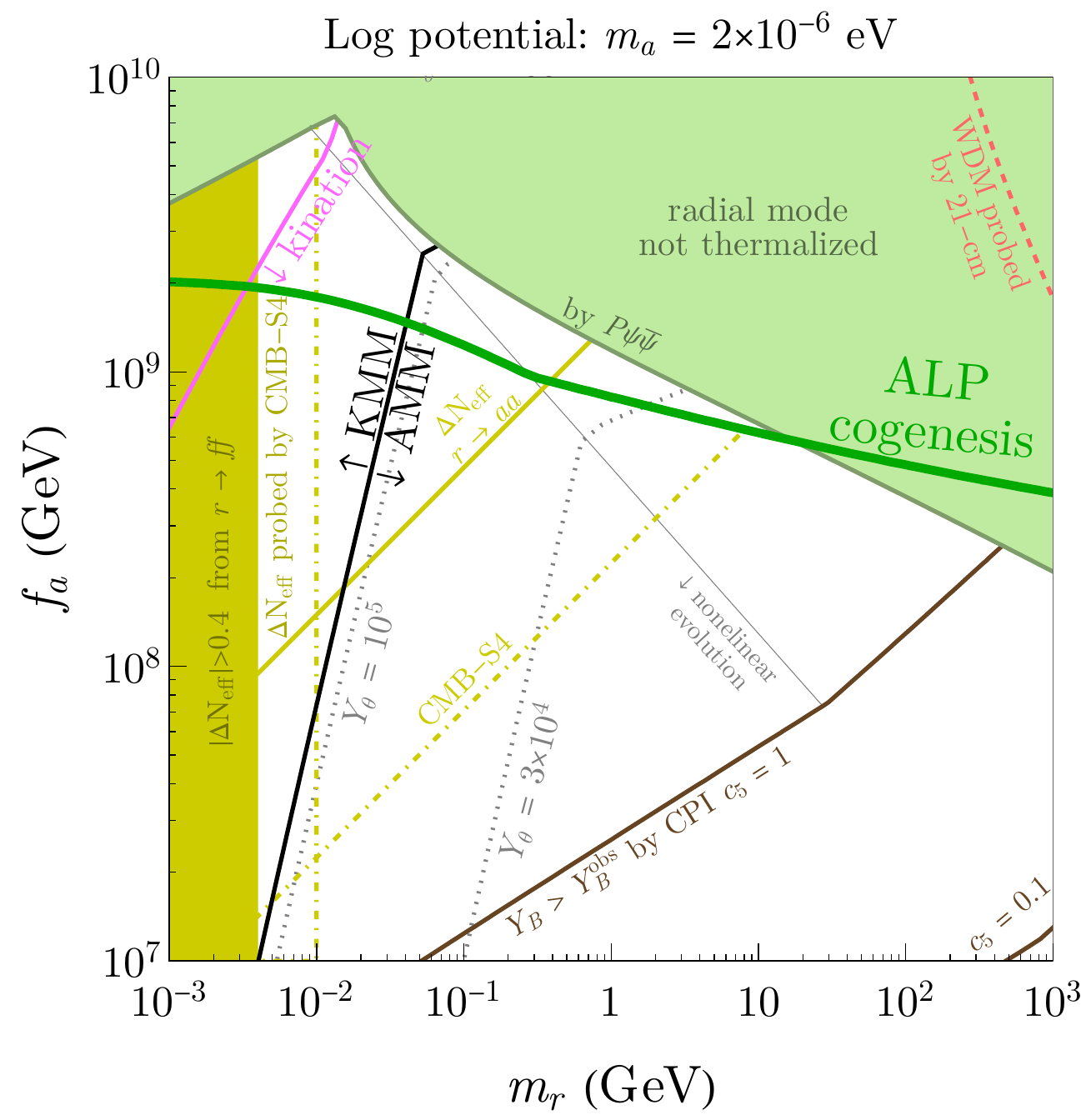}
    \includegraphics[width=0.495\linewidth]{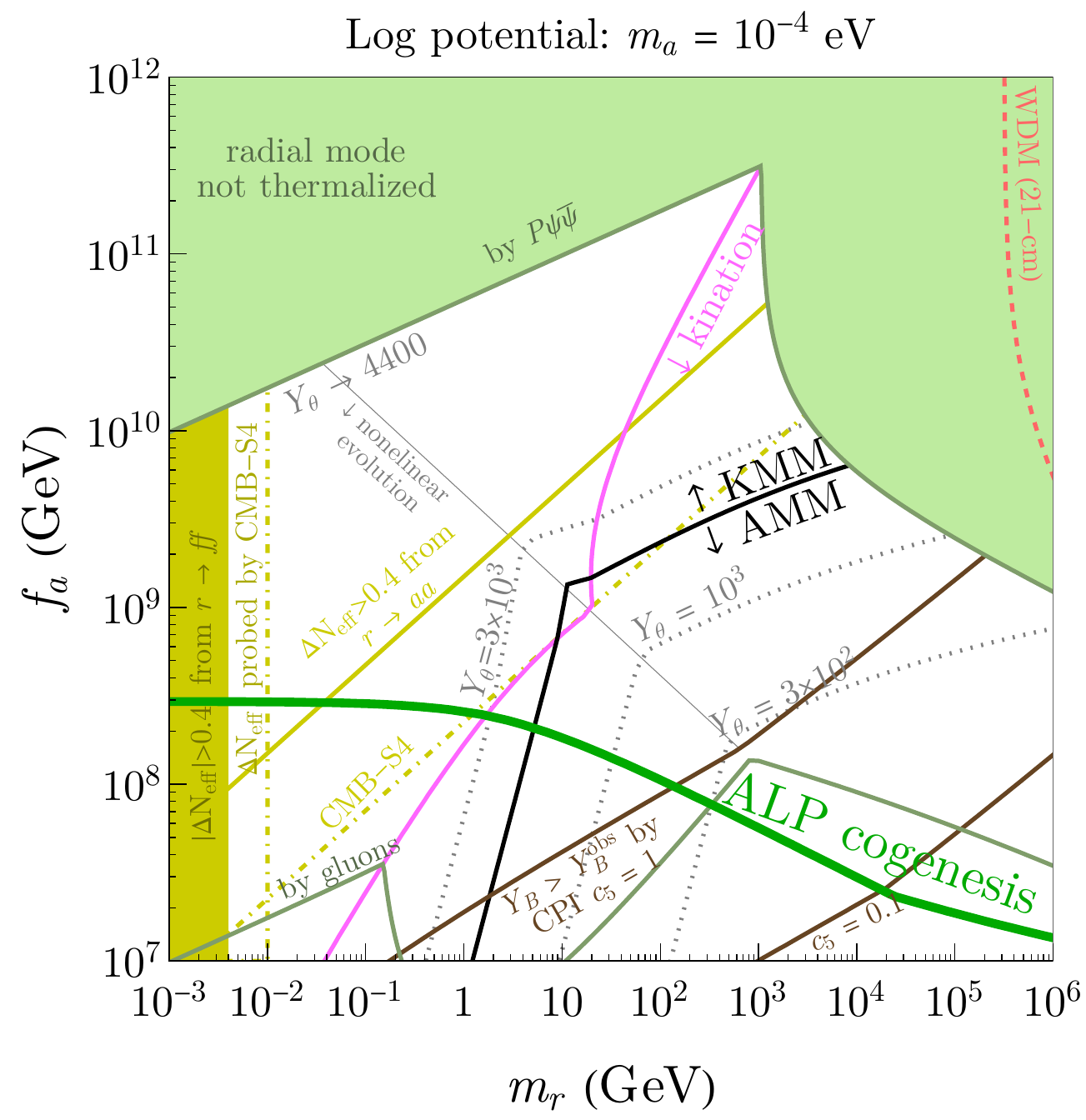}

    \caption{Same as Fig.~\ref{fig:ALP_rP1}, but for the log potential, with $m_a = 2\times10^{-6} \eV$ ($10^{-4} \eV$) in the left (right) panel.}
    \label{fig:ALP_log}
\end{figure}

\begin{figure}[t]
    \centering   
    \includegraphics[width=0.495\linewidth]{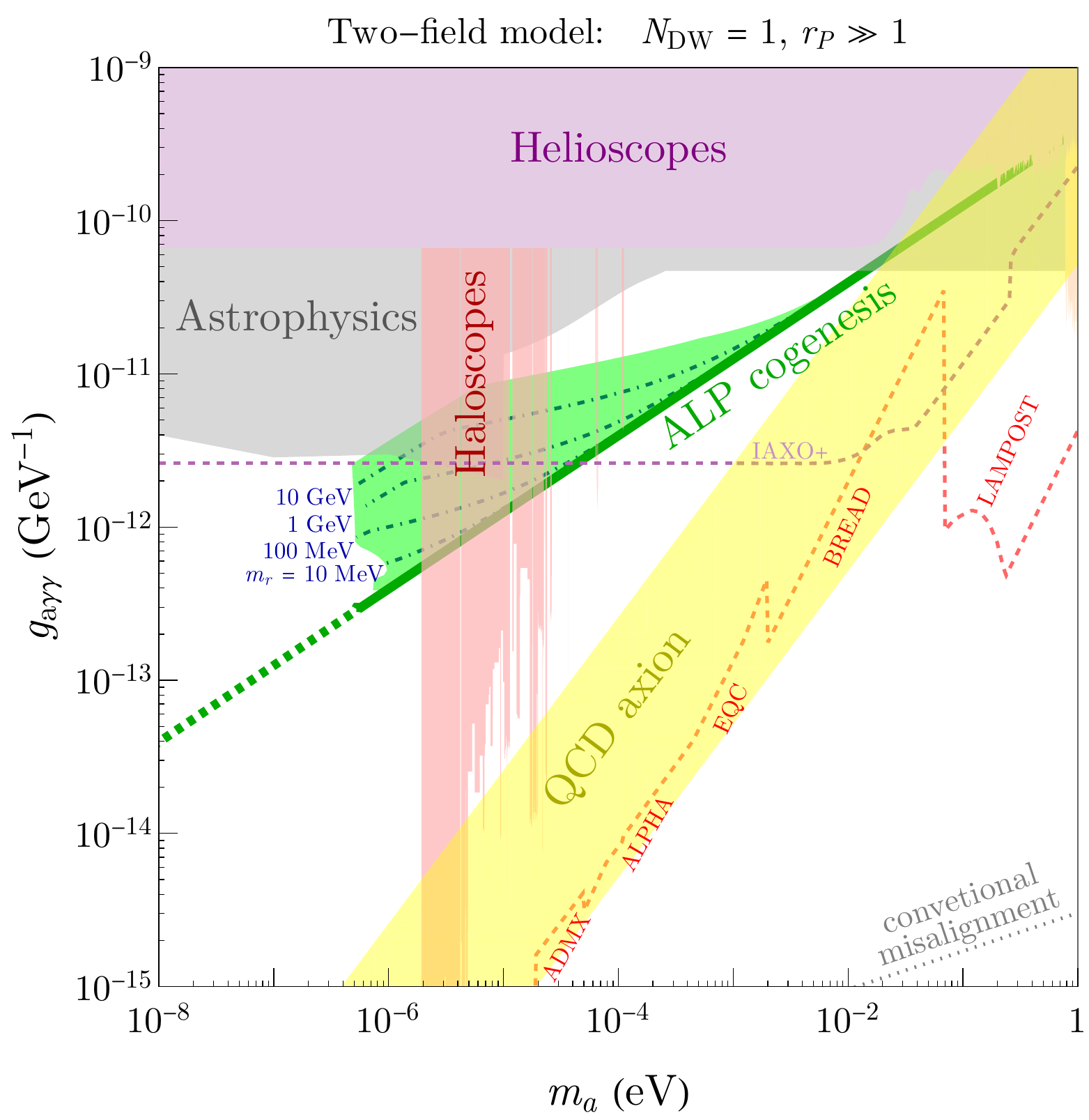}
    \caption{Same as Fig.~\ref{fig:cogenesis}, but for $r_P \gg 1$.}
    \label{fig:cogenesis2}
\end{figure}

In this appendix we show the implications of the AMM for  a broader class of models.
Fig.~\ref{fig:QCDaxion2} is the same as Fig.~\ref{fig:QCDaxion} but for $r_P \gg 1$.
Figs.~\ref{fig:ALP_rPlarge} and \ref{fig:ALP_log} are the same as Fig.~\ref{fig:ALP_rP1} but for $r_P \gg 1$ and the log-potential, respectively. 
Fig.~\ref{fig:cogenesis2} is the same as Fig.~\ref{fig:cogenesis} but for $r_P \gg 1$.

\clearpage

\bibliography{kination} 

\end{document}